\documentclass[prd,aps,showkeys,floats,twocolumn]{revtex4}
\usepackage{amsmath}
\usepackage{amssymb}

\begin{document}

\title{Gravitation and inertia; a rearrangement of vacuum in gravity}

\author{ G.Ter-Kazarian}
\affiliation{Byurakan Astrophysical Observatory, Byurakan, 378433,
Aragatsotn District, Armenia. E-mail: gago\_50@yahoo.com}

\begin{abstract}
We address the gravitation and inertia in the framework of {\em
general gauge} principle (GGP) which accounts for {\em gravitation
gauge group} $G_{R}$ generated by hidden local internal symmetry
implemented on  the flat space. Following the method of
phenomenological Lagrangians, we connect the group $G_{R}$ to
nonlinear realization of the Lie group of {\em distortion} $G_{D}$
of local internal properties of six-dimensional flat space $M_{6},$
which is assumed as a toy model underlying four-dimensional
Minkowski space. We study geometrical structure of the space of
parameters and derive the Maurer-Cartan's structure equations. We
treat distortion fields as Goldstone fields, to which the metric and
connection are related, and infer the group invariants and calculate
conserved currents. The agreement between proposed gravitational
theory and available observational verifications is satisfactory.
Unlike the GR, this theory is free of fictitious forces, which
prompts us  to address separately the inertia from a novel view
point. We construct relativistic field theory of inertia, which
treats inertia as distortion of local internal properties of flat
space $M_{2}$ conducted under the {\em distortion inertial fields}.
We derive the {\em relativistic law of inertia} (RLI) and calculate
inertial force acting on the photon in gravitating system. In spite
of totally different and independent physical sources of gravitation
and inertia, the RLI furnishes justification for introduction of the
Principle of Equivalence. Particular attention is given to
realization of the group $G_{R}$ by the hidden local internal
symmetry of abelian group $U^{loc}=U(1)_{Y}\times diag[SU(2)]$
implemented on the space $M_{6}$. This group has two generators,
third component $T^{3}$ of isospin and hypercharge $Y$ implying
$Q^{d}=T^{3}+Y/2 ,$ where $Q^{d}$ is the {\em distortion charge}
operator assigning the number -1 to particles, but +1 to
anti-particles. This entails two neutral gauge bosons that coupled
to $T^{3}$ and $Y$. We address the rearrangement of vacuum state in
gravity resulting from these ideas. The neutral complex Higgs scalar
breaks the vacuum symmetry leaving the gravitation subgroup intact.
The resulting massive distortion field component may cause an
additional change of properties of spacetime continuum at huge
energies above the threshold value.
\end{abstract}



\maketitle

\section{Introduction}
More than four centuries passed since the famous far-reaching
discovery of Galileo (in $1602-1604$) that all bodies fall at the
same rate~\citep{Drake}, which led to an early empirical version of
suggestion that gravitation and inertia may somehow result from a
single mechanism. Besides describing these early gravitational
experiments, Newton in {\em Principia Mathematica}~\citep{Newt} has
proposed a comprehensive approach to studying the relation between
the gravitational and inertial masses of a body. Ever since there is
an ongoing quest to understand the reason for the equality of
gravitational and inertial forces, which remains an intractable
mystery. From its historical development this can be regarded as
furnishing immediate support for the Principle of Equivalence
asserted by Einstein for General Relativity (GR), which preserves
the idea of relativity of all kinds of motion. Currently, the
Earth-Moon-Sun system provides the best solar system arena for
testing the Principle of Equivalence, for a review see
e.g.~\citep{will0, Tur1}. Any theory of gravitation might explain
both the attraction of masses and inertia in consistent terms.
However, a nature of the relationship of gravity and inertia
continues to elude us and, beyond the Principle of Equivalence,
there has been little progress in discovering their true relation.
Moreover, it seemed that the inertia cannot be ultimately identified
with the gravity as it is proposed in GR, because there are
important reasons to question the validity of this description.
Actually, there are several empirical effects that seem
incomprehensible in this framework. The experiments by~\citep{Coc,
Hug, Drev} tested the important question of anisotropy of inertia
stemming from the idea that the matter in our galaxy is not
distributed isotropically with respect to the earth, and hence if
the inertia is due to gravitational interactions then the inertial
mass of a body will depend on the direction of its acceleration with
respect to the direction towards the center of our galaxy. However,
these experiments do not found such anisotropy of mass. For example,
the most sensitive test is obtained in~\citep{Hug} from a nuclear
magnetic resonance experiment, where the increase in sensitivity
over that which one could obtain from M$\ddot{o}$ssbaur effect is
due to the far narrower line width obtainable for a transition with
a $Li^{7}$ nucleus of spin $I=3/2$ in its ground state as compared
with a nucleus in an excited state. If the mass anisotropy effect is
present, there will be three different intervals which will lead to
a triplet nuclear resonance line, if the structure is resolved, or
to a single broadened line if the structure is unresolved. The
magnetic field was of about $4700$ gauss. The south direction in the
horizontal plane points within $22$ degrees towards the center of
our galaxy, and 12 hour later this same direction along the earth's
horizontal plane points 104 degrees away from the galactic center.
If the nuclear structure of $Li^{7}$ is treated as a single
$P_{3/2}$ proton in a central nuclear potential, the variation
$\Delta m$ of mass with direction, if it exists, was found to
satisfy $\frac{\Delta m}{m}\leq 10^{-20}.$ This proves that there is
no anisotropy of mass which is due to the effects of mass in our
galaxy. If the curvature of Riemannian space is associated with
gravitational interaction then it would indicate an universal
feature equally suitable for action on all the matter fields at
once. Then another objection is that this is rather applicable only
for gravity but not for inertia since the inertia depends solely on
the state of motion of individual test particle or coordinate frame
of interest. So, the curvature arisen due to acceleration of
coordinate frame of interest relates to this coordinate system
itself and does not acts at once on all the other systems or matter
fields. Such interesting aspects which deserve further
investigations, unfortunately, have attracted little attention in
subsequent developments. This state of affairs has not much changed
up to present, as well as the RLI still remain unknown. The present
paper aims to fill this gap. Furnishing justification for
introduction of the Principle of Equivalence, in addition to those
of available experimental verifications~\citep{will0, Tur1}, we must
also assign a high importance to the prove on the theoretical basis.

Another purpose of present article is to explore the rearrangement
of vacuum state in gravity at huge energies, which will be of vital
interest for the physics of superdense matter in very compact
astrophysical sources~\citep{Ter1, Ter2,Ter3,Ter4} and references
therein. Note that the Riemannian space interacting quantum field
theory cannot be a satisfactory ground for addressing this question.
The geometrical interpretation of gravitation arisen from the dual
character of the metrical tensor in its metrical and gravitational
aspects, is a noteworthy result of GR. Although this interpretation
has advantage in solving the problems of cosmology, nevertheless
such a distinction of gravitational field among the fields yields
the difficulties in the unified theories of all interactions of
elementary particles, and in quantization of gravitation. Therefore,
the GR as a geometrized theory of gravitation clashes from the very
outset with basic principles of field theory. This rather stems from
the fact that Riemannian geometry, in general, does not admit a
group of isometries, i.e., Poincar\'{e} transformations no longer
act as isometries and, for example, it is impossible to define
energy-momentum as Noether local currents related to exact
symmetries. This posed severe problems in Riemannian space
interacting quantum field theory. The major unsolved problem is the
non-uniqueness of the physical vacuum and associated Fock space.
Actually, a peculiar shortcoming of the interacting quantum field
theory in curved spacetime is the following two key questions to be
addressed yet: a) an absence of the definitive concept of space-like
separated points, particularly, in canonical approach, and the
'light-cone' structure at each spacetime point; b) the separation of
positive- and negative-frequencies for completeness of the
Hilbert-space description. Due to it, a definition of positive
frequency modes cannot, in general, be unambiguously fixed in the
past and future which leads to $|in>\neq|out>,$ because the state
$|in>$ is unstable against decay into many particle $|out>$ states
due to interaction processes allowed by lack of Poincar\'{e}
invariance. Non-trivial Bogolubov transformation between past and
future positive frequency modes implies that particles are created
from the vacuum and this is one of the reasons for $|in>\neq|out>$.
Note that a remarkable surge of activity of investigations towards
an extension of GR  has arisen recently. They are expressible
geometrically in the language of fundamental structure known as a
fiber bundle. This provides an unified picture of gravity modified
models based on several Lie groups, see e.g.~\citep{BO, Hehl2,
Mansouri1, Mansouri2, Grignani, Chang1, Chang2, StelleWest, IvOg,
IvanovNiederle1, IvanenkoSardanashvily, Lord2, Tresguerres, MAG,
stefano3, felix, symplectism, stefano4}. All these approaches have
their own advantages, but in the same time they are subject to many
uncertainties. Currently no single theory has been uniquely accepted
as the convincing gauge theory of gravitation, which will be able
successfully to address the aforementioned problems.

\subsection{Rational}

To innovate the solution to the problems involved, in this paper we
develop on the {\em general gauge} principle (GGP), an early version
of which is given in ~\citep{Ter4, Ter5, Ter6}. The GGP accounts for
{\em gravitation gauge group} $G_{R}$ generated by hidden local
internal symmetries implemented on the flat space $M_{6}$. Involving
the auxiliary flat space $M_{6}$, with the whole set of well-defined
Killing's vectors, just has a single aim as a guiding tool in
dealing with an intricate 'jungle' of curved geometry. In this paper
much more will be done (Sect.2) to make clear and rigorous these
early results and formulations. The following part of the present
paper will be the original contribution, whereas we relate the group
$G_{R}$ to the Lie group $G_{D}$ of {\em distortion} of local
internal properties of flat space $M_{6}$. It can be achieved by
nonlinear realization of the  group $G_{D}$ in the framework of
method of phenomenological Lagrangians. This approach  was
originally introduced by  Coleman,  Wess and Zumino~\citep{Col, Cal,
W70} in the context of internal symmetry groups. It was later
extended to the case of spacetime symmetries by Isham, Salam, and
Strathdee ~\citep{Isham, Salam}
considering the nonlinear action of $GL(4$, $\mathbf{%
\mathbb{R}
})$ mod the Lorentz subgroup, see~\citep{stefano4} and references
therein for a comprehensive review. We study geometrical structure
of the space of parameters in terms of Cartan's calculus of exterior
forms and derive the Maurer-Cartan's structure equations. We derive
key relation which uniquely determines, for given distortion field,
the six angles of distortion rotations around each axes of the
$M_{6}$. We treat distortion fields as Goldstone fields to which the
metric and connection are related. We infer group invariants and
calculate conserved currents. The metric is no more a fundamental
dynamical field. The fundamental field is distortion gauge field
and, thus, both the actions and the equations of motion depend on
the concept of gauge potential. The metric and connection may be
derived from this gauge field. To test the proposed gravitation
theory, we derive the line element in particular case of static and
spherically symmetric gravitational field. Traditionally the solar
system is a laboratory that offers many opportunities to improve
tests of relativistic gravity. The usual Eddington-Robertson-Schiff
parameters $\beta$ and $\gamma$ used to describe these tests are
perhaps in some sense the most important parameters of the
parameterized post-Newtonian (PPN)
formalism~\citep{will1,will2,Nord,will3}. We rather show that the
agreement is satisfactory between the proposed gravitation theory
and available observational verifications~\citep{shapiro64, will1,
Reas, Rob1, Rob2, Leb, Eub, Shap, Wil1, Wil2, Wil4, Wil5,bbgg92,
will0, Bert, Ies, Tur2}. We complete the proposed theoretical basis
of distortion of spacetime by exploring, further, two major problems
of inertia and rearrangement of vacuum state in gravity.

We construct the relativistic field theory of inertia which
similarly to gravitation theory treats the inertia effects  as a
distortion of local internal properties of flat spacetime continuum.
We motivate this approach as follows. Unlike the GR, proposed gauge
field theory of gravitation is free of fictitious forces, because
the infinite-parameter group of general covariance is no longer in
use. Instead, the preferred systems and group of transformations of
the, so-called, {\em real-curvilinear coordinates} relate solely to
real gravitational fields. In spite of totally different and
independent physical sources of gravitation and inertia, still we
might expect that the inertial force is of the same nature as
gravitational force. Namely we ascribe the effects associated with
gravity and inertia to spacetime geometry itself, and that both
phenomena arise due to the distortion of local internal properties
of flat space. To trace this line,  we involve besides the
distortion gauge fields being responsible only for gravitation, also
the {\em distortion inertial fields} which account for the inertia
separately. Seeking a replacement for the unobservable Newtonian
{\em absolute spacetime}, which is necessary to assign a meaning to
Newtonian {\em absolute acceleration}, instead we explore the
geometry of two-dimensional flat space $M_{2}$. Similar reasoning
leads us, further, to the conclusion that an alteration of uniform
motion of test particle under the unbalanced force is the immediate
cause of the real distortion of the local internal properties of the
space $M_{2} $ conducted under the distortion inertial field. This
necessarily, in the first place, with equal justice could be
interpreted as a definite criterion for the universal {\em absolute}
acceleration  of test particle or coordinate frame of interest, and
in the second place, will give us the fundamental RLI. This we might
expect to hold on the basis of an intuition founded on a past
experience limited to low velocities, and which were implicit in the
ideas of Galileo and Newton as to the nature of inertia. The major
premise is that the centrifugal endeavor of particles to recede from
the axis of rotation is directly proportional to the quantity of the
{\em absolute} circular acceleration, which, for example, concave
water surface in Newton's famous rotating bucket experiments. In
this framework, the {\em relative} acceleration (in Newton's
terminology) (both magnitude and direction), in contrary, cannot be
the cause of the distortion of the space $M_{2}$ and, thus, it does
not produce inertia effect. Therefore, the real {\em inertia}
effects can be an empirical indicator of {\em absolute}
acceleration. We calculate the inertial force acting on the photon
in gravitating system  of particles that are bound together by their
mutual gravitational attraction. A particular attention is given to
the theoretical justification for introduction of the Principle of
Equivalence.

Finally, the developments on the GGP  are applied to address the
rearrangement of vacuum state in gravity. The objections concerning
non-uniqueness of the physical vacuum can be circumvented
 immediately due to one of the underlying
principles that in the flat space interacting quantum field theory
the vacuum is well-determined and unique $|in>=|out>$ (up to a phase
factor). In realization of $G_{R}$  we implement the simplest hidden
gauge symmetry of abelian group $U^{loc}=U(1)_{Y}\times diag[SU(2)]$
on the $M_{6}$ which entails two neutral gauge bosons. Spontaneous
symmetry breaking is achieved in standard manner by introducing the
neutral complex Higgs scalar. Non-vanishing vacuum expectation value
(VEV) leaves one Goldstone boson which is gauged away from the
scalar sector. But it essentially reappears in the gauge sector
providing the longitudinally polarized spin state of one of gauge
bosons that acquires mass through its coupling to Higgs scalar. The
massless component of distortion field is responsible for
gravitational interactions. In the resulting theory, simultaneously
with the strong gravity, the massive distortion  field component may
cause a substantial change of properties of spacetime continuum at
huge energies above the threshold value.

This paper is organized as follows: In Sect.2 a number of useful
mathematical concepts of GGP are reviewed for the reader's
convenience. We will refrain from providing lengthy details of the
formalism of GGP and unitary map. For these the reader is referred
to Appendix. In Sect.3 we relate the group $G_{R}$ to the Lie group
$G_{D}$ by constructing its nonlinear realization. In Sect.4 we
construct the relativistic field theory of inertia and give
theoretical justification for introduction of the Principle of
Equivalence. In Sect.5 we address the rearrangement of vacuum state
in gravity. Conclusions are presented in Sect.6. The specific topics
dealt with in the Appendix are further details on the GGP. We will
be brief and often suppress the indices without notice. Unless
otherwise stated we take natural units, $h=c=1$. The quantities
denoted by wiggles throughout this paper refer to distorted (curved)
space, but the quantities referring to flat space are left without
wiggles.

\section{The GGP preliminaries}
For the benefit of the reader, a brief outline of the framework of
GGP are given in this Section and in Appendix to make the rest of
the paper understandable. We have used a combined geometrical
structure known as a fiber bundle, which provides a unified picture
of theory based on the local internal gauge symmetries. The gravity,
as a gauge theory, could be achieved by introducing a generalized
gauge transformation law (Eq.~(\ref{R1})) which enables the gauging
of external spacetime groups.

Given the principal fiber bundle $\widetilde{\mathbb{P}}(
R_{4}\text{, }G_{R}; \,\widetilde{\pi} )$  with the structure group
$G_{R}$,  the local coordinates
$\widetilde{p}\in\widetilde{\mathbb{P}}$ are $ \widetilde{p}=(
\widetilde{x}\text{, }U_{R}), $ where $\widetilde{x}\in R_{4}$ and
$U_{R}\in G_{R},$ the total bundle space $\widetilde{\mathbb{P}}$ is
smooth manifold, the surjection $\widetilde{\pi} $ is a smooth map
$\widetilde{\pi} :\widetilde{\mathbb{P}}\rightarrow R_{4}.$  The
base space is assumed to be curved four dimensional Riemannian space
$R_{4}$ in order to describe the effects of gravitation. A set of
open coverings $\{ \widetilde{\mathcal{U}}_{i}\} $ of $R_{4}$ with
$\widetilde{x}\in \{ \widetilde{\mathcal{U}}_{i}\} \subset R_{4}$
satisfy $\bigcup\nolimits_{\alpha } \widetilde{\mathcal{U}}_{\alpha
}=R_{4}.$  The fibration is given as
$\bigcup\nolimits_{\widetilde{x}} \widetilde{\pi} ^{-1}(
\widetilde{x}) =\widetilde{\mathbb{P}}.$ The local gauge will be the
diffeomorphism map $\widetilde{\chi}
_{i}:\widetilde{\mathcal{U}}_{i}\times _{R_{4}}G_{R} \rightarrow
\widetilde{\pi}{} ^{-1}(\widetilde{\mathcal{U}} _{i})\in
\widetilde{\mathbb{P}}, $ since $\widetilde{\chi} _{i}^{-1}$ maps
$\widetilde{\pi}{} ^{-1}(\widetilde{\mathcal{U}}_{i})$\ onto the
direct (Cartesian) product $\widetilde{\mathcal{U}_{i}}\times
_{R_{4}}G_{R}$. Here $\times _{R_{4}}$ represents the fiber product
of elements defined over space $ R_{4} $ such that $\widetilde{\pi}
( \widetilde{\chi} _{i}\left( \widetilde{x}\text{, }U_{R}\right) )
=\widetilde{x}$ and $ \widetilde{\chi} _{i}( \widetilde{x}\text{,
}U_{R}) =\widetilde{\chi} _{i}( \widetilde{x}\text{, }( id)
_{G_{R}}) U_{R}=\widetilde{\chi} _{i}( \widetilde{x}) U_{R}\ \forall
\widetilde{x}\in \{ \widetilde{\mathcal{U}}_{i}\},$ $( id) _{G_{R}}$
is the identity element of group $G_{R}$. Let the collection of
matter fields of arbitrary spins $\widetilde{\Phi}(\widetilde{x})$
(the various suffixes are left implicit) take values in standard
fiber over
$\widetilde{x}:\,\widetilde{\pi}{}^{-1}(\widetilde{\mathcal{U}}_{i})
=\widetilde{\mathcal{U}}_{i}\times
\widetilde{\mathbb{F}}_{\widetilde{x}}.$ The fiber
$\widetilde{\pi}{} ^{-1}$ at $\widetilde{x}\in R_{4}$  is
diffeomorphic to $\widetilde{\mathbb{F}},$ where
$\widetilde{\mathbb{F}}$ is the fiber space, such that $
\widetilde{\pi}{} ^{-1}\left( \widetilde{x}\right) \equiv
\widetilde{\mathbb{F}}_{\widetilde{x}}\approx
\widetilde{\mathbb{F}}.$ The action of structure group $G_{R}$ on
$\widetilde{\mathbb{P}}$ defines an isomorphism of the Lie algebra
$\widetilde{\mathfrak{g}}$ of $G_{R}$ onto the Lie algebra of
vertical vector fields on $\widetilde{\mathbb{P}}$ tangent to the
fiber at each $\widetilde{p}\in \widetilde{\mathbb{P}}$ called
fundamental. Whereas, the tangent and cotangent bundles,
respectively, are $\widetilde{T}( \widetilde{\mathbb{P}})$ and
$\widetilde{T}{}^{\ast }( \widetilde{\mathbb{P}}) $ ,
$\widetilde{T}_{p}( \widetilde{\mathbb{P}}) $ is the space of
tangents at $\widetilde{p}\in \widetilde{\mathbb{P}}$, i.e.
$\widetilde{T}_{p}( \widetilde{\mathbb{P}}) \in \widetilde{T}(
\widetilde{\mathbb{P}}) $.
  The metric is the section of conjugate vector bundle
$S^{2}\widetilde{T}{}^{\ast }( \widetilde{\mathbb{P}})$ (symmetric
part of tensor degree): $\,\hat{g}:\widetilde{T}(
\widetilde{\mathbb{P}}) \times \widetilde{T}(
\widetilde{\mathbb{P}}) \rightarrow C^{\infty} (R_{4}),$ where a
section is a smooth map $S:R_{4}\rightarrow
\widetilde{\mathbb{P}}\text{,} $ such that $S(\widetilde{x})\in
\widetilde{\pi}{} ^{-1}(\widetilde{x})
=\widetilde{\mathbb{F}}_{\widetilde{x}} \forall \widetilde{x}\in
R_{4},$ and satisfies $\pi \circ S=(id) _{R_{4}}, $ where $(\circ )$
represents the group composition operation, where $( id) _{R_{4}}$
is the identity\ element of $R_{4}$. It assigns to each point
$\widetilde{x}\in R_{4}$ a point in the fiber over $\widetilde{x}$.
The general coordinate transformations $\delta
\widetilde{x}{}=f(\widetilde{x})$, where $f(\widetilde{x})$ is an
arbitrary function of coordinates $\widetilde{x}$, yield the
infinite-parameter group of general covariance in $R_{4}$ if only
the functions $f(\widetilde{x})$ can be expanded in power series of
$\widetilde{x}$. The expansion coefficients are considered as the
group-parameters, and that the group-algebra includes an infinite
number of generators.

{\em Remark:} An invariance of the Lagrangian $L_{\widetilde{\Phi}}$
of matter fields $\widetilde{\Phi}(\widetilde{x})$  under the
infinite-parameter group of general covariance in $R_{4}$ implies an
invariance of $L_{\widetilde{\Phi}}$ under the {\em gravitation
gauge group} $G_{R}$ and vice versa if, and only if, the generalized
local gauge transformations of the fields
$\widetilde{\Phi}(\widetilde{x})$ and their covariant derivative
$\nabla_{\mu}\widetilde{\Phi}(\widetilde{x})$ are introduced by
finite local $U_{R}(\in G_{R})$ gauge transformations as
\begin{equation}
\begin{array}{l}
\widetilde{\Phi}'(\widetilde{x})=U_{R}\,(\widetilde{x})
\,\widetilde{\Phi}(\widetilde{x}),  \\
\left[ g^{\mu}(\widetilde{x})\,\nabla_{\mu}\widetilde{\Phi}
(\widetilde{x})\right]' =U_{R}\,(\widetilde{x}) \left[
g^{\mu}(\widetilde{x})\,\nabla_{\mu}\widetilde{\Phi}
(\widetilde{x})\right],
\end{array}
\label{R1}
\end{equation}
where $\nabla_{\mu}$ denotes the covariant derivative agreed with
the metric, $\,g^{\mu}(\widetilde{x})\rightarrow
\widetilde{e}{}^{\mu}(\widetilde{x})$ for the fields of spin
$(j=0,1)$,  and $g^{\mu}(\widetilde{x})=
V^{\mu}_{\alpha}(\widetilde{x})\,\gamma^{\alpha}$  for the spinor
field $(j=\frac{1}{2})$, where
$V^{\mu}_{\alpha}(\widetilde{x})=<\widetilde{e}{}^{\mu},\,\widetilde{e}_{\alpha}>$
are the components of affine tetrad vectors
$\widetilde{e}{}^{\alpha}$ in used coordinate net
$\widetilde{x}{}^{\mu}$~\citep{Bir}, $\gamma^{\alpha}$ are the
Dirac's matrices. The unitary matrix  $U_{R}\,(\widetilde{x})$ will
be determined below.

Next, suppose the massless gauge field $ a(x)\,(\equiv a_{\mu}(x))$
takes  values in Lie algebra $\mathfrak{g}$ of abelian group
$U^{loc},$ which is a local form of expression of connection in
principle fiber bundle $\mathbb{P}( M_{4}\text{, }U^{loc};\,\pi) $
with the structure group $U^{loc}$ and the surjection $\pi$. The
base space is the flat Minkowski space $ M_{4},$ so, a set of open
coverings $\{ \mathcal{U}_{i}\} $ of $M_{4}$ with $x\in \{
\mathcal{U}_{i}\} \subset M_{4}$ satisfy $\bigcup\nolimits_{\alpha }
\mathcal{U}_{\alpha }=M_{4}.$ The metric is the section of conjugate
vector bundle  $\hat{\eta}:T( \mathbb{P}) \times T( \mathbb{P})
\rightarrow C^{\infty} (M_{4}),$ whereas the symmetric components
$(\eta_{l k})$ of metrical tensor can be given in basis $(e_{l})$.
The matter fields $\Phi(x)$ of arbitrary spin are the sections of
vector bundles associated with abelian group $U^{loc}$. They take
values in standard fiber which is the Hilbert vector space where a
linear representation $U(x)$ of group $U^{loc}$ is given. This space
can be regarded as Lie algebra of group $U^{loc}$ upon which Lie
algebra acts according to law of adjoint representation: $
a\,\leftrightarrow \, ad \, a\: \, \Phi \, \rightarrow [a\, ,\Phi].
$  We adopt the following conventions: Greek indices stand for
variables in $R_{4}$, Latin indices refer to $M_{4}$, and that
$\psi^{\mu}_{l}\equiv
\partial_{l}\, \widetilde{x}{}^{\mu}$ where
$\partial_{l}=\partial/\partial \,x^{l}$. Aforesaid is the
mathematical tools of conventional gauge dynamics. Now, to involve a
drastic revision of a role of gauge fields in physical concept of
curved geometry, below we generalize this scheme by exploring a new
special type of {\em distortion} gauge fields assumed acting on
external spacetime groups. While, a local internal gauge symmetry
$U^{loc}(1)$ remains hidden symmetry as far as it is screened by the
gravitation gauge group $G_{R}$.

{\em Theorem 1:} For any generalized gauge field dynamics of
Eqs.~(\ref{R1}) defined on $R_{4}$ the underlying (surjective)
conventional gauge field dynamics can always be constructed on
$M_{4}$.

{\em Proof:} The following three steps are the very foundation of
our construction procedure which went into the proof of this
theorem.

{\em First step:} We assume that the basis vector $(e)$ undergoes
{\em distortion} transformations under the {\em distortion} gauge
field $(a)$:
\begin{equation}
\begin{array}{l}
\widetilde{e}{}_{\mu}(a)= D_{\mu}^{l}(a)\,e_{l}. \label{RE2}
\end{array}
\end{equation}
The transformation matrix $D(a)$ will be determined in Sect.3.

{\em Second step:} We construct the diffeomorphism\\
$\widetilde{x}{}^{\mu}(x,a):M_{4}\rightarrow R_{4}$ by seeking the
new holonomic coordinates $\widetilde{x}{}^{\mu}(x,a)$ as the
solutions of the first-order partial differential equations
\begin{equation}
\begin{array}{l}
\widetilde{e}_{\mu}(a)\,\psi^{\mu}_{l}=\Omega^{m}_{l}(F)e_{m},
\end{array}
\label{RE3}
\end{equation}
where $\Omega^{m}_{l}(F)=\delta^{m}_{l}+\omega^{m}_{l}(F),$ the
$(F)$ denotes antisymmetrical tensor of gauge field
$F_{nk}=\partial_{n}\,a_{k}-\partial_{k}\,a_{n}$, and hence the
tensor $\Omega^{m}_{l}(F)$ has a null variational derivative $
(\delta \Omega^{m}_{l}(F)/\delta a_{n}) = 0 $ at local variations of
connection $a_{n} \rightarrow a_{n}+ \delta a_{n}.$

{\em Third step:} We consider a smooth unitary map of all the matter
fields and their covariant derivatives:
\begin{equation}
\begin{array}{l}
R(a):\Phi\rightarrow \widetilde{\Phi},\\
S(a)\,R(a):\left(\gamma^{k}D_{k}\Phi\right)\rightarrow
\left(g^{\nu}(x)\nabla_{\nu}\widetilde{\Phi}\right),
\end{array}
\label{R2}
\end{equation}
where $R(a)$  is the unitary map matrix, $S(F)$ is the gauge
invariant scalar function (see Eq.~(\ref{R9}) and App.2),
$D_{k}=\partial_{k}- i\ae\,a_{k}$, $\ae$ is the gauge coupling
constant which relates to Newton gravitational constant as in
Eq.~(\ref{R19}).

The conditions of integrability $
\partial_{k}\,\psi^{\mu}_{l}=\partial_{l}\,\psi^{\mu}_{k}
$ and non-degeneracy ($\| \psi \| \neq 0$) necessarily
hold~\citep{Dub, Pont}, therefore, the following constrain is
imposed upon the tensor $\Omega^{m}_{l}(F)$: $
\partial_{k} ( D^{\mu}_{m}\Omega^{m}_{l})=\partial_{l} (
D^{\mu}_{m}\Omega^{m}_{k}), \label{RE2E} $ the solution of which can
be written in general form $ \Omega^{m}_{l}(F)=
D_{\nu}^{m}(a)\partial_{l} \Theta^{\nu}(a,F), $ where
$\Theta^{\mu}(a,F)$ are the arbitrary functions such that
$\partial_{k}
\partial_{l}\Theta^{\mu}(a,F)=\partial_{l}
\partial_{k}\Theta^{\mu}(a,F).$ Hence, the equation~(\ref{RE3}) yields
the bilinear form $d\,\widetilde{s}{}^{2}$ on $R_{4}:$
\begin{equation}
\begin{array}{l}
d\,\widetilde{s}{}^{2}=g_{\mu\nu}\,d\,\widetilde{x}{}^{\mu}d\,\widetilde{x}{}^{\nu}=
ds^{2}_{\chi}\equiv\\\Omega^{m}_{l}(F)\Omega^{m}_{k}(F)dx^{l}dx^{k}=inv(\Lambda,\,
U^{loc}),
\end{array}
\label{RE6}
\end{equation}
where  $ds^{2}_{\chi}$ is the Lorentz ($\Lambda$) and gauge
($U^{loc}$) invariant line element given on $M_{4}$. Denoting
$\chi_{l}=\omega^{m}_{l}(F)e_{m},$ and
$\chi^{\mu}_{l}=\psi^{\mu}_{l}-D^{\mu}_{l},$ we may derive the
following gauge invariant scalar functions:
\begin{equation}
\begin{array}{l}
\chi(F)=\,<e^{l},\,\chi_{l}(F)>=\omega^{l}_{l}(F)=tr\,\omega(F),\\
S(F)= \frac{1}{4}\psi_{\mu}^{l}(a,F)D^{\mu}_{l}(a)=
1+\frac{1}{4}tr\,\omega(F).
\end{array}
\label{R9}
\end{equation}
In what follows,  we take the form
\begin{equation}
\begin{array}{l}
\omega^{m}_{l}(F)=\delta^{m}_{l}\omega(x)(F),
\end{array}
\label{REE77}
\end{equation}
where we do not initially specify the scalar function $\omega(F)$
apart that $\omega(0)=0.$ Instead, at some intermediate stage in the
analysis we adopt an expansion form (Subsect.3.3). The curvature of
the space $R_{4}$  is zero if $ (\partial
\psi_{\mu}^{l}/d\widetilde{x}{}^{\nu})=\Gamma^{\lambda}_{\mu\nu}\,
\psi_{\lambda}^{l} $~\citep{W72}, where $\Gamma^{\lambda}_{\mu\nu}$
denote the Christoffel symbols agreed with the metric
$g_{\mu\nu}(a)=D_{\mu}^{l}(a)D_{\nu}^{l}(a).$ In illustration of the
point at issue, the Eqs.~(\ref{R2}) explicitly may read
\begin{equation}
\begin{array}{l}
\widetilde{\Phi}{}^{\mu\cdots
\delta}(\widetilde{x})=\psi^{\mu}_{l}\cdots
\psi^{\delta}_{m}\,R(a)\,\Phi^{l\cdots m}(x)\equiv
\\{(R_{\psi})}^{\mu \cdots \delta}_{l\cdots m}\,\Phi^{l\cdots m}(x),
\end{array}
\label{R3}
\end{equation}
and that
\begin{equation}
\begin{array}{l}
g^{\nu}(x)\nabla_{\nu}\widetilde{\Phi}{}^{\mu\cdots
\delta}(\widetilde{x})=\\S(F)\, \psi^{\mu}_{l}\cdots
\psi^{\delta}_{m}\,R(a)\,\gamma^{k}D_{k} \,\Phi^{l\cdots m}(x).
\end{array}
\label{R4}
\end{equation}
Using the gauge transformations in $M_{4}$, it is a straightforward
to determine from Eq.~(\ref{R3}) the matrix $U_{R}\,(\widetilde{x})$
in terms of matrices $U$ and $R$. Actually,
$$\widetilde{\Phi}'(\widetilde{x})=U_{R}\,(\widetilde{x})
\,\widetilde{\Phi}(\widetilde{x})=
U_{R}\,R_{\psi}(a)\,\Phi=R_{\psi}'\,\Phi'=R_{\psi}'\,U\,\Phi.
$$ Similarly, this can be  determined from Eq.~(\ref{R4}) too:
$$
\begin{array}{l}
\left(g^{\nu}(x)\nabla_{\nu}\widetilde{\Phi}(\widetilde{x})\right)'=U_{R}\,(\widetilde{x})
\,\left(g^{\nu}(x)\nabla_{\nu}\widetilde{\Phi}(\widetilde{x})\right)=\\
U_{R}\,S(F)\,R_{\psi}(a)\,\left(\gamma^{k}D_{k}\Phi\right)=
S(F')\,R_{\psi}'\,\left(\gamma^{k}D_{k}\,\Phi\right)'=\\
S(F')\,R_{\psi}'\,U\,\left(\gamma^{k}D_{k}\,\Phi\right).
\end{array}
$$
Hence $ U_{R}= R_{\psi}'\,U\,R_{\psi}^{-1}, $ where $R_{\psi}'\equiv
R_{\psi}(a')$ and the $(a')$ denotes $U^{loc}$-transformed gauge
field. Based on the Theorem 1 we may extend conventional gauge
principle to involve gravity in the GGP scheme by requiring that:
{\em The physical system of the fields
$\widetilde{\Phi}(\widetilde{x})$ defined on $R_{4}$ must always be
invariant under the finite local gauge transformations $U_{R}$ of
the Lie group of gravitation $G_{R}.$ }
\begin{center}
\begin{picture}(70,60)
\put(-80,30){$\widetilde{\Phi}'(\widetilde{x})= U_{R}\,
\widetilde{\Phi}(\widetilde{x})$}
\put(10,40){\shortstack{$U_{R}=R_{\psi}'\,U\,R_{\psi}^{-1}$}}
\put(95,35){\vector(-1,0){90}}
\put(110,30){$\widetilde{\Phi}(\widetilde{x})$}
\put(100,1){\vector(0,1){30}}
\put(110,10){$R_{\psi}(\widetilde{x},\,x)$} \put(110,-10){$\Phi(x)$}
\put(45,-1){\shortstack{U}} \put(95,-5){\vector(-1,0){90}}
\put(-80,-10){$\Phi'(x)=U \,\Phi(x)$} \put(0,1){\vector(0,1){30}}
\put(-55,10){$R_{\psi}'(\widetilde{x},\,x)$}
\end{picture}
\end{center}
\vskip 0.5truecm
\begin{center} The scheme of GGP.
\end{center}
Although, in the reminder of this article we have explored the
simplest abelian symmetry $U^{loc}$ as hidden symmetry, however, one
may envisage that a straightforward extension should be to achieve
the full machinery of the GGP scheme for non-abelian symmetries. We
conclude, on the observations above that out of all the arbitrary
coordinates in $R_{4}$ the {\em real-curvilinear} coordinates
$\widetilde{x}(x,a)$ can be distinguished which are derived from
Eq.~(\ref{RE3}) at all Lorentz ($\Lambda$) and gauge ($U^{loc}$)
transformations of variables $(x)$ and $(a)$. Hence, unlike GR, the
wider infinite-parameter group of general covariance in $R_{4}$ is
no longer in use. Therefore, some Lorentz or gauge transformation
necessarily underlies the arbitrary transformation $\widetilde{x}
\rightarrow \widetilde{x}{}'$ of real-curvilinear coordinates which
relate solely to real gravitational fields. This prompts us to treat
the inertia separately (Sect.4). In case of zero curvature, the
Eq.~(\ref{RE3}) can be satisfied globally in $M_{4}$ by setting $
\psi^{\mu}_{l}=D^{\mu}_{l}=V^{\mu}_{l}=(\partial x^{\mu}/\partial
X^{l}), \quad \|D\|\neq 0,\quad \chi_{l}=0, $ where $X^{l}$ are the
inertial coordinates. In this,  one has conventional gauge theory
given on the $M_{4}$  in both curvilinear and inertial coordinates.
At this point, we have discussed all the mathematical tools that
complete the formalism of GGP by making clear and rigorous the early
results and formulations~\citep{Ter4, Ter5, Ter6}. In what follows
we shall present the original contribution.

\section{Nonlinear realization of distortion group $G_{D}$}
The nonlinear realization  technique~\citep{Col, Cal} provides a way
to determine the transformation properties of fields defined on the
quotient space $G/H$. Constructing nonlinear realization of the Lie
group of distortion $G_{D}$, first, within the scheme of the GGP we
necessarily introduce the language of a conceptual six-dimensional
geometry of $M_{6}$, which is assumed as a toy model underlying the
$M_{4}$. This replacement appears to be indispensable in discussion
of  the distortion of local internal properties of spacetime
continuum, and it mostly manifests its virtues in constructing the
relativistic field theory of inertia (Sect.4). So, let  $M_{6}$ be
the smooth differentiable six-dimensional flat space with the
decomposition law as follows:
\begin{equation}
\begin{array}{l}
M_{6}=R^{3}_{+}\oplus R^{3}_{-}=R^{3}\oplus T^{3},
\\sgn(R^{3})=(+++), \quad sgn(T^{3})=(---).
\end{array}
\label{R31}
\end{equation}
The $e_{(\lambda\alpha)}=O_{\lambda}\times \sigma_{\alpha}\quad
(\lambda = \pm, \, \alpha=1,2,3)$ are linear independent unit basis
vectors at the point (p) of interest of given three-dimensional
space $R^{3}_{\lambda}$. The unit vectors $O_{\lambda}$ and $
\sigma_{\alpha}$ imply
\begin{equation}
\begin{array}{l}
<O_{\lambda},\,O_{\tau}>={}^{*}\delta_{\lambda\tau}, \quad
<\sigma_{\alpha},\,\sigma_{\beta}>=\delta_{\alpha\beta},
\end{array}
\label{R32}
\end{equation}
where $\delta_{\alpha\beta}$ is the Kronecker symbol, and
${}^{*}\delta_{\lambda\tau}=1-\delta_{\lambda\tau}.$  Consequently,
three spatial $e_{\alpha}=\xi\times \sigma_{\alpha}$ and three
temporal $e_{0\alpha}=\xi_{0}\times \sigma_{\alpha}$ components are
the basis vectors, respectively, in spaces $R^{3}$ and $T^{3}$,
where $O_{\pm}=(1/\sqrt{2})(\xi_{0}\pm \xi),\quad
\,\xi^{2}_{0}=-\xi^{2}=1, \quad <\xi_{0},\,\xi>=0.$ Within this
scheme, we are presumably allowed to perceive directly the
three-dimensional ordinary space $R^{3}$, but not the three -
dimensional time space $T^{3}$ being orthogonal to the former. In
using this language it is important to guard a reduction to the
space $M_{4}$ which can be achieved in the
following way. \\
1) In case of free flat space $M_{6}$, the subspace $T^{3}$ is
isotropic. And so far it contributes in line element just only by
the square of the moduli  $t=|\textbf{x}^{0}|,\, \textbf{x}^{0}\in
T^{3}$, then, the reduction $M_{6} \rightarrow M_{4}=R^{3} \oplus
T^{1}$ can be readily achieved if for conventional {\em time} we use
$t=|\textbf{x}^{0}|$.
\\
2) In case of curved space, the reduction $R_{6}\rightarrow R_{4}$
can be achieved if we use the projection ($\widetilde{e}_{0}$) of
the temporal component ($\widetilde{e}_{0\alpha} $) of basis
six-vector
$\widetilde{e}\,\,(\widetilde{e}_{\alpha},\,\widetilde{e}_{0\alpha})$
on the given {\em universal} direction ($\widetilde{e}_{0\alpha}
\rightarrow \widetilde{e}_{0}$) . By this we choose the {\em time}
coordinate. Actually, the Lagrangian of physical fields defined on
$R_{6}$ is the function of scalars such as $A_{(\lambda
\alpha)}B^{(\lambda \alpha)}= A_{\alpha}B^{\alpha} + A_{0
\alpha}B^{0 \alpha}$, then upon the reduction of temporal components
of six-vectors $A_{0 \alpha}B^{0 \alpha}=A^{0
\alpha}<\widetilde{e}_{0\alpha}, \widetilde{e}_{0\beta}>B^{0
\beta}=A^{0}<\widetilde{e}_{0}, \widetilde{e}_{0}>B^{0}=A_{0}B^{0} $
we may fulfill a reduction to $R_{4}$.

\subsection{Distortion of local internal properties of the $M_{6}$}
First, we consider distortion transformations of the ingredient unit
vectors $O_{\tau}$ under the distortion gauge field $(a)$:
\begin{equation}
\begin{array}{l}
\widetilde{O}_{(+\alpha)}(a)={\cal Q}^{\tau}_{(+\alpha)}(a)\,
O_{\tau}=O_{+}+\ae \, a_{(+\alpha)}O_{-},\\
\widetilde{O}_{(-\alpha)}(a)={\cal Q}^{\tau}_{(-\alpha)}(a)\,
O_{\tau}=O_{-}+\ae \, a_{(-\alpha)}O_{+},
\end{array}
\label{R35}
\end{equation}
where ${\cal Q}\,\left(={\cal Q}_{(\lambda\alpha)}^{\tau}(a)\right)$
is the element of the group $Q$. This violates the first relation in
Eq.~(\ref{R32}) because of $\widetilde{O}_{(\lambda\alpha)}^{2}(a)$
$=2\ae \, a_{(\lambda\alpha)}\neq 0$ for given $\lambda$ and
$\alpha$. Next, we assume that this induces the distortion
transformations of ingredient unit vectors $\sigma_{\beta}$, which,
in turn, undergo the rotations: $
\widetilde{\sigma}_{(\lambda\alpha)}(\theta)={\cal
R}^{\beta}_{(\lambda\alpha)}(\theta)\, \sigma_{\beta}, $ where
${\cal R}(\theta)\in SO(3)$ is the element of the group of rotations
of planes involving two arbitrary axes around orthogonal third axis
in the given ingredient space $R^{3}_{\lambda}$. Then, resulting
basis vectors $ \widetilde{\sigma}_{(\lambda\alpha)}(\theta)$ of
each three-dimensional ingredient space $R^{3}_{\lambda}$  retain
the orthogonality condition between themselves, but violate it
between the basis vectors of different ingredient spaces. That is,
$<\widetilde{\sigma}_{(\lambda\alpha)},\,\widetilde{\sigma}_{(\tau\beta)}>_{\alpha\neq\beta}\neq
0,  \quad\mbox{at}\quad \lambda\neq\tau.$ In fact, distortion
transformations of basis vectors ($O$) and ($\sigma$) are not
independent, and  rather governed by the spontaneous breaking of
distortion symmetry (see Eq.~(\ref{R646})). To avoid a further
proliferation of indices, hereafter we will use upper case Latin
$(A)$ in indexing $(\lambda\alpha)$, etc. The infinitesimal
transformations then read
\begin{equation}
\begin{array}{l}
\delta\,{\cal Q}_{A}^{\tau}(a) = \ae \, \delta\,a_{A}
X^{\tau}_{\lambda}\,\in Q, \\ \delta\,{\cal
R}(\theta)=-\frac{i}{2}M_{\alpha\beta}\delta\,\omega^{{\alpha\beta}}\,\in
SO(3),
\end{array}
\label{R636}
\end{equation}
provided by generators $X^{\tau}_{\lambda}=
{}^{*}\delta^{\tau}_{\lambda}$ and $I_{i}=\frac{\sigma_{i}}{2}$,
where $\sigma_{i}$ are the Pauli's matrices,  $
M_{\alpha\beta}=\varepsilon_{\alpha\beta\gamma}I_{\gamma}$ and $
\delta\,\omega^{{\alpha\beta}}=\varepsilon_{\alpha\beta\gamma}\delta\,\theta_{\gamma}.$
Transformation matrix $ D(a,\,\theta)= {\cal Q}(a)\times {\cal
R}(\theta) $ is the element of distortion group $G_{D}=Q\times
SO(3)$:
\begin{equation}
\begin{array}{l}
D_{(d\,a^{A},\,d\,\theta^{A})}= I+ d\,D_{(a^{A},\,\theta^{A})},\\
d\,D_{(a^{A},\,\theta^{A})}=
i\left[d\,a^{A}X_{A}+d\,\theta^{A}I_{A}\right],
\end{array}
\label{R638}
\end{equation}
where $I_{A}\equiv I_{\alpha}$ at given $\lambda$. We may join to
each point $(a,\,\theta)$ of group space the Descartes' reper which
is equal, in group sense, to the reper joint to point of origin
$(0,\,0)$. This is introduced to ensure that the vector
$(a,\,\theta; \,a+d\,a,\,\theta+d\,\theta)$ has the same analytical
expression of the vector $(0,\,0; \,d\,a',\,d\,\theta)'$:
(Eq.~(\ref{R638}))
\begin{equation}
\begin{array}{l}
D_{(a,\,\theta)}\,d\,D^{-1}_{(a,\,\theta)}=
i[\omega^{A}\,X_{A}+\vartheta^{A}\,I_{A}],
\end{array}
\label{R838}
\end{equation}
where we denote $d\,a^{'A}=\omega^{A}(a,\,\theta;
\,d\,a,\,d\,\theta)$ and $d\,\theta^{'A}=\vartheta^{A}(a,\,\theta;\,
d\,a,\,d\,\theta)$. Then,
\begin{equation}
\begin{array}{l}
d\,e_{A}= e_{A}\,d\,F_{A}=O_{A}(d)\times \sigma_{A}+O_{A}\times
\sigma_{A}(d)=\\
i[\omega^{A}(d)\,X_{A}+\vartheta^{A}(d)\,I_{A}]\,e_{A},
\end{array}
\label{R738}
\end{equation}
where $e_{A}\equiv(\exp F_{A})$. The functions $\omega^{A}(d)$ and
$\vartheta^{A}(d)$ will be determined in the next subsection.

\subsection{A spontaneous breaking of distortion symmetry}
Following the method of phenomenological Lagrangians \citep{Col,
Cal, W70, BO, Og, stefano4} and references therein, our goal is to
treat the distortion group $G_{D}$ and its stationary subgroup
$H=SO(3)$, respectively, as dynamical group and its algebraic
subgroup. But the generators $X_{A}$ (Eq.~(\ref{R636})) of group Q
do not complete the group H to the dynamical group $G_{D}$,
therefore, they cannot be interpreted as the generators of quotien
space $G_{D}/H$, and that the distortion fields $a_{A}$ cannot be
identified directly with the Goldstone fields arisen in spontaneous
breaking of distortion symmetry $G_{D}$. These objections, however,
can be circumvented if we define the pair of the original basis
vectors $O_{\lambda}$ in terms of orthogonal unit vectors $\xi_{0}$
and $\xi$. The distortion transformations Eq.~(\ref{R35}) of basis
vectors $O_{\pm}$ then immediately become rotations of the group
$SO(3)$ of modified basis vectors
\begin{equation}
\begin{array}{l}
\overline{\widetilde{O}}_{A}=\frac{1}{\sqrt{2}}
\left(\widetilde{\xi}_{0A}(\overline{\theta})+\epsilon_{A}\,\widetilde{\xi}_{A}(\overline{\theta})\right)
=\cos\,\overline{\theta}_{A}\widetilde{O}_{A},
\end{array}
\label{R634}
\end{equation}
where $\tan\,\overline{\theta}_{A}\equiv-\ae \, a_{A},\,$
$\epsilon_{(+\alpha)}=-\epsilon_{(-\alpha)}=1$,  and that
\begin{equation}
\begin{array}{l}
\left(
  \begin{array}{c}
    \widetilde{\xi}_{0A}(\overline{\theta}) \\
    \widetilde{\xi}_{A}(\overline{\theta}) \\
  \end{array}
\right) = \left(
  \begin{array}{cc}
    \cos\,\overline{\theta}_{A} & \sin\,\overline{\theta}_{A} \\
    -\sin\,\overline{\theta}_{A} & \cos\,\overline{\theta}_{A} \\
  \end{array}
\right) \left(
  \begin{array}{c}
    \xi_{0} \\
    \xi \\
  \end{array}
\right).
\end{array}
\label{R633}
\end{equation}
Here, a rotation on $\overline{\theta}_{(+\alpha)}$ is clockwise,
while on $\overline{\theta}_{(-\alpha)}$ is counterclockwise.
Consequently, the distortion group $G_{D}=Q\times SO(3)$ can be
mapped in one-to-one manner onto the group $G_{D}=SO(3)\times SO(3)$
which, in turn, is isomorphic to chiral group $SU(2)\times SU(2)$.
In this case the method of phenomenological Lagrangians is well
known. For a convenience, throughout this subsection we leave the
Greek indices implicit unless otherwise stated: $A=(\lambda
i)\rightarrow i=1,2,3,$. But it goes without saying that all the
results obtained refer to the given $R^{3}_{\lambda}$ space. Three
$I_{i}$ among six generators of the group correspond to isotopic
transformations, and three $K_{i}$- to special chiral
transformations mixing the states of different parities. They imply
the conventional commutation relations
\begin{equation}
\begin{array}{l}
\left[I_{i},\,I_{j}\right]=i\varepsilon_{ijk} I_{k}, \quad
\left[I_{i},\,K_{j}\right]=i\varepsilon_{ijl} K_{l},  \\
\left[K_{i},\,K_{j}\right]=i\varepsilon_{ijk} I_{k},
\end{array}
\label{R640}
\end{equation}
of invariant subgroup $H=SO(3)$, with the generators $I_{i}$, and of
quotien space $G_{D}/H$, with the generators $K_{i}$, where
$\varepsilon_{ijk}$ denotes the antisymmetric unit tensor. Three
modified parameters $\overline{a}{}^{i}(a)$ of the quotien space
$G_{D}/H$ of adjacent classes, with respect to which the Lagrangian
of physical fields is not invariant, can be identified with three
Goldstone fields. They are introduced to make provisions for the
Eq.~(\ref{R738}), which incorporated into Eq.~(\ref{R634}) yields
\begin{equation}
\begin{array}{l}
d\,\overline{e}_{A}(\overline{a})=\overline{e}_{A}(a)\,d\,\overline{F}_{A}(\overline{\theta},\,\theta)=
\overline{O}_{A}(d)\times \sigma_{A}+\\\overline{O}_{A}\times
\sigma_{A}(d)= i[\omega^{i}(\overline{a},\,d\,\overline{a})\,K_{i}+
\vartheta^{l}(\overline{a},\,d\,\overline{a})\,I_{l}]\,\overline{e}_{A}(\overline{a}).
\end{array}
\label{R739}
\end{equation}
This is written in terms of generators of the group $SU(2)\times
SU(2)$, where
$\overline{e}_{A}(\overline{a})=\overline{O}_{A}(\overline{a})\times
\sigma_{A}(\overline{a})$. We are at once led to seek the function
$\overline{\theta}(\theta)$ if
$d\,\overline{F}_{A}(\overline{\theta},\,\theta)$ constitutes a
total differential, i.e.,
$\omega^{i}(\overline{a},\,d\,\overline{a})$ and
$\vartheta^{l}(\overline{a},\,d\,\overline{a})$ are  Cartan's forms.
This implies
\begin{equation}
\begin{array}{l}
\frac{\partial^{2}\,\overline{F}_{A}}{\partial\,
\overline{\theta}\,\partial\,\theta}=
\frac{\partial^{2}\,\overline{F}_{A}}{\partial\,\theta\,\partial\,
\overline{\theta}},
\end{array}
\label{R1740}
\end{equation}
where $\overline{\theta}\equiv \overline{\theta}_{A}$ and
$\theta\equiv\theta_{A}$. Using the infinitesimal transformations
(see Eq.~(\ref{R636}))
\begin{equation}
\begin{array}{l}
d\,\sigma_{(\lambda l)}= \sigma_{(\lambda
l)}(d)=\frac{1}{2}\varepsilon_{lkj}\sigma_{k}d\,\theta_{(\lambda
j)},
\end{array}
\label{R648}
\end{equation}
it is straightforward to calculate the partial derivative
\begin{equation}
\begin{array}{l}
\frac{\partial\,F_{A}}{\partial\, \theta}=
\frac{\partial\,\sigma_{A}}{\sigma_{A}\,\partial\, \theta}\equiv
\frac{\sigma_{A}(\partial)}{\sigma_{A}\,\partial\, \theta}=
\frac{1}{\sin\, \theta_{A}}.
\end{array}
\label{R649}
\end{equation}
The similar relation  holds for the vectors
$d\,\widetilde{\xi}_{0A}(\overline{\theta})$ and
$d\,\widetilde{\xi}_{A}(\overline{\theta})$ (Eq.~(\ref{R633})), and
that $(\partial\,F_{A}/\partial\, \overline{\theta})=(1/\sin\,
\overline{\theta}_{A})$. Upon the reduction, the holonomy condition
Eq.~(\ref{R1740}) becomes $ (\partial/\partial\,
\theta)\left(1/\sin\,
\overline{\theta}_{A}\right)=(\partial/\partial\,
\overline{\theta})\left(1/\sin\, \theta_{A}\right), $ with
nontrivial solution $\theta_{A}=\overline{\theta}_{A}$. Hence we
arrive at
\begin{equation}
\begin{array}{l}
\tan\,\theta_{A}=-\ae\, a_{A}.
\end{array}
\label{R646}
\end{equation}
Given distortion field $a_{A}$, the key relation~(\ref{R646})
uniquely determines six angles $\theta_{A}$ of rotations around each
of six ($A$) axes. We are now in position to derive the
Maurer-Cartan's structure equations. According to Poincar\'{e}'s
theorem,  the exterior derivative $(')$ of total differential form
is zero, and that the Eq.~(\ref{R739}) gives
\begin{equation}
\begin{array}{l}
(\overline{e}_{A}(a)\,d\,\overline{F}_{A})'=\overline{e}_{A}(a)\,\left([d\,\overline{F}_{A},
\delta\,\overline{F}_{A}]+
(d\,\overline{F}_{A})'\right)=\\i\left([\omega^{i}(\overline{a},\,d\,\overline{a})\,K_{i}+
\vartheta^{l}(\overline{a},\,d\,\overline{a})\,I_{l}]\,\overline{e}_{A}(a)\right)'=0,
\end{array}
\label{R740}
\end{equation}
where $[d\,\overline{F}_{A},
\delta\,\overline{F}_{A}]=(d\,\overline{F}_{A})' =0.$ In calculating
of exterior differential and exterior product of the forms, in
Eq.~(\ref{R740}) the differentials of functions
$(K_{i}\,\overline{e}_{A})$ and $(I_{l}\,\overline{e}_{A})$ figured
in bilinear differential
\begin{equation}
\begin{array}{l}
\delta\,d\,\overline{e}_{A}=
i[\delta\,\omega^{i}(d)\,X_{i}+\delta\,\vartheta^{l}(d)\,I_{l}]\,\overline{e}_{A}+\\
i[\omega^{i}(d)\,\delta(K_{i}\,\overline{e}_{A})+\vartheta^{l}(d)\,\delta(I_{l}\,\overline{e}_{A})],
\end{array}
\label{R751}
\end{equation}
are defined according to Eq.~(\ref{R739}). Equating to zero the
coefficients at the same linearly independent generators of exterior
differential in Eq.~(\ref{R740}), this yields the following system
of equations:
\begin{equation}
\begin{array}{l}
(\omega^{i})'=[\omega^{k},\,\vartheta^{\beta}]\,\varepsilon_{ik\beta},\\
(\vartheta^{\gamma})'=[\vartheta^{\alpha},\,\vartheta^{\beta}]\,\varepsilon_{\gamma\alpha\beta}/2+
\varepsilon_{\gamma k i}[\omega^{k},\,\omega^{i}]/2.
\end{array}
\label{RU7}
\end{equation}
Defining new forms
$\omega^{i}_{k}=\vartheta^{\beta}\varepsilon_{ik\beta}$, and using
the relations $ \varepsilon_{k\alpha\beta}\,\varepsilon_{i j
k}=(\varepsilon_{i\beta k}\,\varepsilon_{k\alpha
j}-\varepsilon_{i\alpha k}\,\varepsilon_{k\beta j})$, which stem
from the Yacobi's identity
\begin{equation}
\begin{array}{l}
\{ [K_{i}[I_{\alpha},\,I_{\beta}]]\}+\quad\mbox{permutations}\equiv
0,
\end{array}
\label{RU8}
\end{equation}
the Eqs.~(\ref{RU7}) yield the Maurer-Cartan's structure equations
\begin{equation}
\begin{array}{l}
(\omega^{i})'=[\omega^{k},\,\omega^{i}_{k}],\\
(\omega^{i}_{k})'=-R^{l}_{jki}[\omega^{k},\,\omega^{i}]/2
+[\omega^{k}_{j},\,\omega^{l}_{k}],
\end{array}
\label{R647}
\end{equation}
where $R^{l}_{jki}=-\varepsilon_{lj\gamma}\varepsilon_{\gamma ki}$
is the curvature of the group space. These equations, as usual,
describe the motion of orthogonal reper joint to given point of
group space. The forms $\omega^{i}$ and $\omega^{i}_{k}$ are
interpreted as transformations of translation and rotation of the
orthogonal reper, i.e., the rotation $(\vartheta^{l}I_{l})$ belongs
to stationary group $H$, but the translation reads
$(\omega^{i}K_{i})$. The invariant constraint Eq.~(\ref{R646}) has
as an immediate consequence that there always exists the rotation
transformation of stationary subgroup $H$ annulling the change in
equation of Cartan's forms arisen from the translation
transformation of the quotien space $\overline{Q}=G_{D}/H$. The
forms $\omega^{i}$ and $\vartheta^{l}$ can be used to construct the
group invariants, namely the phenomenological Lagrangians. The
Lagrangian of Goldstone fields ($\overline{a}$) can be identified
with the square of interval of the geodesic line, with minimal
number of derivatives, between the infinitely closed points
$\overline{a}{}^{i}$ and $\overline{a}{}^{i}+d\,\overline{a}{}^{i}$:
\begin{equation}
\begin{array}{l}
L_{\overline{a}}(\eta)=\frac{1}{2}\,\omega^{i}(\overline{a},\,\partial_{A}\,\overline{a})\,
\omega^{i}(\overline{a},\,\partial_{A}\,\overline{a}),
\end{array}
\label{R840}
\end{equation}
where $\varepsilon_{\alpha i l}\varepsilon_{l\alpha j}=-\delta_{ij}$
is the metrical tensor of the group space,
$\partial_{A}=(\partial/\partial \eta^{A})$,  $\eta$ is the local
coordinates in open neighborhood of $p\in M_{6}$. In normal
coordinates it becomes
\begin{equation}
\begin{array}{l}
L_{\overline{a}}(\eta)=(\partial_{A}\,\overline{a})^{2}/2 +\\
(\delta_{ik}-\overline{a}{}^{i}\overline{a}{}^{k}/\overline{a}{}^{2})\,
\left(
\sin^{2}\,\sqrt{\overline{a}{}^{2}}/\sqrt{\overline{a}{}^{2}}-1\right)
\,\partial_{A}\,\overline{a}{}^{i}\,\partial_{A}\,\overline{a}{}^{k}/2.
\end{array}
\label{R999}
\end{equation}
Since the massless gauge field $ (a)$ associates with the gauge
group $U^{loc}$, the Lagrangian Eq.~(\ref{R999}) should be equated
to undegenerated Killing form defined on the Lie algebra of the
group $U^{loc}$ in adjoint representation
\begin{equation}
\begin{array}{l}
L_{\overline{a}}(\eta)= L_{a}(\eta)= -\frac{1}{4}<F_{AB}(a),\,
F^{AB}(a)>_{K},
\end{array}
\label{R841}
\end{equation}
where  $F_{AB}(a)$ is the antisymmetrical tensor of gauge field
$(a)$. The Goldstone fields ($\overline{a}$) can be determined  from
the Eq.~(\ref{R841}) as the functions of gauge field $(a)$. The
covariant derivatives of matter fields $\Phi$ interacting with the
Goldstone fields ($\overline{a}$) can be determined by means of the
form $\vartheta^{\alpha}$ as
\begin{equation}
\begin{array}{l}
L=L_{0}\left(\Phi,\,\partial_{A}\Phi+\vartheta^{\alpha}(\overline{a},\,
\partial_{A}\,\overline{a})T_{\alpha}\,\Phi\right),
\end{array}
\label{R842}
\end{equation}
where $L_{0}\left(\Phi,\,\partial_{A}\Phi\right)$ is the Lagrangian
of interacting matter fields classified by the linear
representations $T_{\alpha}$ of the subgroup $H$. The Lagrangians
(\ref{R841}) and  (\ref{R842}) are invariant with respect to
distortion translations and rotations. This must be completed by the
transformations of the fields $\Phi$
\begin{equation}
\begin{array}{l}
\Phi'=\left(\exp \left[i
\theta'{}^{\alpha}(\overline{a},\,g)T_{\alpha}\right]\right)\Phi,
\end{array}
\label{Rgg3}
\end{equation}
where $\theta'{}^{\alpha}(\overline{a},\,g)$ is given by
Eq.~(\ref{R844}).

\subsection{Static line element with spherical-symmetry}
The field equations can be derived from an invariant action $
S=S_{a} + \widetilde{S}_{\widetilde{\Phi}}$, which is similar to
Eq.~(\ref{R15}). The action of distortion gauge field $S_{a}$ given
on the flat space $M_{6}$ is invariant under the Lorentz ($\Lambda$)
and gauge ($U^{loc}$) groups, while the action of matter fields
$\widetilde{S}_{\widetilde{\Phi}}$ given on the curved space $R_{6}$
is invariant under the gauge group of gravitation $G_{R}$. Field
equations may immediately follow in terms of Euler-Lagrange
variations carried out in the spaces $M_{6}$ and $R_{6}$,
respectively. The field equation for dimensionless potential
$x_{A}\equiv\ae\,a_{A}$ can be obtained from the Lagrangian
(\ref{R841}) as
\begin{equation}
\begin{array}{l}
\partial^{B}
\partial_{B}\,
x_{A} - (1-\zeta^{-1}_{0})\partial
_{A}\partial^{B}\,x_{B}=\\-\frac{1}{2}\,\ae^{2}\,\sqrt{g(x)}\,\frac{\partial
g ^{BC}(x)}{\partial x_{A}} \widetilde{T}_{BC},
\end{array}
\label{R401}
\end{equation}
where $\partial_{B}=\partial/\partial \eta^{B}$, $\eta^{B}$ are the
coordinates in the given space $R^{3}_{B},$  $\widetilde{T}_{BC}$
denotes the energy-momentum tensor, $\,\zeta_{0}$ is the gauge
fixing parameter. To render our discussion here more transparent,
below we clarify a relation between gravitational and coupling
constants. To assist in obtaining actual solutions from field
equations, we may consider the weak-field limit and shall envisage
that the right hand side of Eq.~(\ref{R401}) should be in the form
\begin{equation}
\begin{array}{l}
-\frac{1}{2}\,(4\pi G_{N})\,\sqrt{g(x)}\,\frac{\partial g
^{BC}(x)}{\partial x_{A}} \widetilde{T}_{BC}.
\end{array}
\label{R402}
\end{equation}
Hence, we may assign to the Newton's gravitational constant $G_{N}$
the value
\begin{equation}
\begin{array}{l}
G_{N}=\ae^{2}/4\pi.
\end{array}
\label{R19}
\end{equation}
{\em Remark:} Although the distortion gauge field ($a_{A}$) is
vector field, nevertheless the key relation Eq.~(\ref{R19}) shows
that only the gravitational attraction presents in proposed theory
of gravitation.

To obtain a feeling for this point we may consider physical systems
which are static as well as spherically symmetrical. Upon the
reduction $R_{6}\rightarrow R_{4}$, we have the group of motions
$SO(3)$ with two-dimensional space-like orbits $S^{2}$ where the
standard coordinates are $\widetilde{\theta}$ and
$\widetilde{\varphi}$. The stationary subgroup of $SO(3)$ acts
isotropically upon the tangent space at the point of sphere $S^{2}$
of radius $\widetilde{r}$. So, the bundle $\widetilde{\pi}:R_{4}
\rightarrow \widetilde{R}^{2}$ has the fiber
$S^{2}=\widetilde{\pi}{}^{-1}(\widetilde{x})$, $\quad
\widetilde{x}\in R_{4}$ with a trivial connection on it, where
$\widetilde{R}^{2}$ is the quotient-space $R_{4}/SO(3)$. The
coordinates $\widetilde{x}^{\mu}(\widetilde{t},\widetilde{r},
\widetilde{\theta},\widetilde{\varphi})$ implying the diffeomorphism
$ \widetilde{x}^{\mu}(x, a):M_{4}\rightarrow R_{4} $ exist in the
whole region $\widetilde{\pi}{}^{-1}(\widetilde{\mathcal{U}}_{i})\in
R_{4}$, where $\widetilde{x}^{0r}\equiv \widetilde{t}, \quad
\widetilde{x}^{0\theta}= \widetilde{x}^{0\varphi}=0.$  In outside of
configuration of given mass, the field equation Eq.~(\ref{R401}) can
be written in Feynman gauge as $\nabla^{2}a_{0}=0,$ which has the
solution $x_{0}=- r_{g}/2r,$ where $x_{0}\equiv\ae\,a_{0}(r)$,
$\,r_{g}$ is the gravitational radius. The components of
transformation matrix are $ D^{0}_{\widetilde{0}}=1+x_{0},\quad
D^{r}_{\widetilde{r}}=1-x_{0},\quad
D^{\theta}_{\widetilde{\theta}}=D^{\varphi}_{\widetilde{\varphi}}=1.$
From Eq.~(\ref{RE3}) and Eq.~(\ref{REE77}), we obtain
\begin{equation}
\begin{array}{l}
\frac{\partial \widetilde{x}^{\mu}}{\partial x^{l}}\equiv
\psi^{\mu}_{l}= D^{\mu}_{l}(1+\omega(F)).
\end{array} \label{R42}
\end{equation}
It can be easily verified  that either $\partial_{r}\psi^{0}_{0}\neq
\Gamma^{0}_{01}\psi^{0}_{0},$ or $\partial_{r}\psi^{1}_{1}\neq
\Gamma^{1}_{11}\psi^{1}_{1},$ etc., and the curvature of the space
$R_{4}$ is not zero. The line element then reads
\begin{equation}
\begin{array}{l}
d\,\widetilde{s}{}^{2}=(1-\frac{r_{g}}{2r})^{2}\,d\widetilde{t}^{2}-
(1+\frac{r_{g}}{2r})^{2}\,d\widetilde{r}^{2}-\\\widetilde{r}^{2}
(\sin^{2}\widetilde{\theta} \,d\,\widetilde{\varphi}{}^{2}+
d\,\widetilde{\theta}{}^{2}), \label{R43}
\end{array}
\end{equation}
provided, the Eq.~(\ref{R42}) and  Eq.~(\ref{R43}) give the relation
\begin{equation}
\begin{array}{l}
\frac{dg_{00}}{d r}=\frac{dg_{00}}{d
\widetilde{r}}\frac{1+\omega(F)}{1+\frac{r_{g}}{2r}}=\frac{r_{g}}{r^{2}}(1-\frac{r_{g}}{2r}),
\end{array}
\label{R42E}
\end{equation}
for determining the function $\widetilde{r}(r).$ We must now turn to
the actual correspondence between the  expression~(\ref{R43}) for
the line element surrounding an attracting central body and the
observational facts of astronomy. The investigating methods are so
well known that it will be sufficient for our purposes merely to
indicate the classical tests of GR conducted in solar-system dealing
only with the shape of the trajectories of photons and planets. All
these tests are carried out in empty space and in gravitational
fields that are to a good approximation static and spherically
symmetric. Therefore, in sufficient approximation, at great
distances from the central body
\begin{equation}
\begin{array}{l}
F=(4r_{g}^{2}{\ae}^{2}F_{mn}F_{mn})^{1/4} =r_{g}/r << 1,
\end{array}
\label{RG2}
\end{equation}
we may take expansion of function $ \omega(F)=
\lambda_{1}F+\lambda_{2}F^{2}+\cdots, $ and that from
Eq.~(\ref{R42E}) we obtain
\begin{equation}
\begin{array}{l}\widetilde{r}=r(1+\alpha_{1}F+\alpha_{2}F^{2}+\cdots),
\end{array}
\label{RG3}
\end{equation}
provided $\alpha_{1}=1/2-\lambda_{1},$
$\alpha_{2}=\lambda_{2}+1/4+4(\lambda_{1}-1/4)^{2},$ etc. Then, in
terms up to the second order in
$\widetilde{F}\quad(=r_{g}/\widetilde{r}),$ which is an
approximation of interest for available observational verifications,
the temporal component of metrical tensor reduces to $g_{00}\simeq 1
- \widetilde{F}+(\lambda_{1}-1/4)\widetilde{F}{}^{2}.$ With these
provisions, Eq.~(\ref{R43}) is reduced to standard Schwarzschild
line element with the metrical tensor components as follows:
\begin{equation}
\begin{array}{l}
g_{00}\simeq 1 -
\frac{r_{g}}{\widetilde{r}}+(\lambda_{1}-\frac{1}{4})\frac{r_{g}^{2}}{\widetilde{r}^{2}},\\
g_{11}\simeq -( 1 + \frac{r_{g}}{\widetilde{r}}+\cdots ), \\
g_{22}=-\widetilde{r}{}^{2}, \quad
g_{33}=-\widetilde{r}{}^{2}\,\sin^{2} \widetilde{\theta}.
\label{R44}
\end{array}
\end{equation}
The free adjustable parameter $\lambda_{1}$ in Eq.~(\ref{R44}) can
be written in terms of Eddington-Robertson expansion parameters as
$\lambda_{1}= 1/4+2(\beta-\gamma).$ While $\gamma$ controls also
other relativistic effects, in particular those related to
gravito-magnetism, it mainly affects electromagnetic propagation.
The differential displacement of the stellar images near the Sun
historically was the first experimental effect to be investigated
and is now of great importance in accurate astrometry. The bending
of a light ray also increases the light-time between two points, an
important effect usually named after its discoverer I. I. Shapiro
~\citep{shapiro64}. Several experiments to measure this delay have
been successfully carried out, using \emph{wide-band} microwave
signals passing near the Sun and transponded back, either passively
by planets, or actively, by space probes, see~\citep{will1,Reas}.
The very long baseline interferometry (VLBI) has achieved accuracies
of better than 0.1 mas (milliarcseconds of arc), and regular
geodetic VLBI measurements have frequently been used to determine
the space curvature parameter $\gamma$~\citep{Rob1, Rob2, Leb, Eub,
Shap}, resulting in the accuracy of better than $\sim0.045\%$ in the
tests of gravity via astrometric VLBI observations. Detailed
analysis of VLBI data have yielded a consistent stream of
improvements $\gamma =1.000±0.003 $~\citep{Rob1, Rob2},
$\gamma=0.9996±0.0017$~\citep{Leb},
$\gamma=0.99994±0.00031$~\citep{Eub} and $\gamma
=0.99983±0.00045$~\citep{Shap} resulting in the accuracy of better
than $\sim0.045\%$. The major advances in several disciplines
notably in microwave spacecraft tracking, high precision astrometric
observations, and lunar laser ranging (LLR) suggest new experiments.
LLR, a continuing legacy of the Apollo program, provided improved
constraint on the combination of parameters
$4\beta-\gamma-3$~\citep{Wil1, Wil2, Wil4, Wil5}. The analysis of
LLR data~\citep{Wil2} constrained this combination as
$4\beta-\gamma-3=(4.0 \pm 4.3)\times10^{-4},$ leading to an accuracy
of $\sim0.011\%$ in verification of general relativity via precision
measurements of the lunar orbit. A significant improvement was
reported in 2003 from Doppler tracking of the Cassini spacecraft
while it was on its way to Saturn~\citep{bbgg92}, with a result
$\gamma-1 =(2.1 \pm 2.3)\times10^{-5}.$ This was made possible by
the ability to do Doppler measurements using both X-band (7175 MHz)
and Ka-band (34316 MHz) radar, thereby significantly reducing the
dispersive effects of the solar corona. In addition, the 2002
superior conjunction of Cassini was particularly favorable: With the
spacecraft at 8.43 astronomical units from the Sun, the distance of
closest approach of the radar signals to the Sun was only $1.6
R_{\odot}$. This experiment has reached the current best accuracy of
$\sim0.002\%$~\citep{Bert}. Keeping in mind aforesaid, the best fit
for satisfactory agreement between the proposed theory of
gravitation and observation can be reached at
$\lambda_{1}-1/4=(2.95\pm 3.24)\times10^{-5}$.

\section{The relativistic field theory of inertia}
As we mentioned in Sect.2, in the proposed theory of gravitation,
the preferred systems and group of transformations of the {\em
real-curvilinear} coordinates relate only to the real gravitational
fields. This prompts us to introduce separately the {\em distortion
inertial fields}, which have other physical source than that of
gravitation, and construct the relativistic field theory of inertia.
The latter, similarly to gravitation theory, treats the {\em
inertia} as a distortion of local internal properties of flat
$M_{2}$ space. The geometry of Sect.3 is a language which is almost
indispensable for the treatment of this problem.

\subsection{The case of unbalanced net force other than gravitational }
First,  we shall discuss the inertia effects in particular case when
the relativistic test particle accelerated in the flat space under
unbalanced net force other than gravitational. Let us concentrate
our attention on the first observer in two-dimensional Minkowski
flat space $M_{2}=R_{(+)}^{1}\oplus R_{(-)}^{1}=R^{1}\oplus T^{1}$,
being regarded as in a state of rest or uniform motion. Suppose this
unaccelerated observer for the position  of free test particle in
$M_{2}$ uses the inertial coordinate frame $S_{(2)}$ corresponding
to spatial $q\in R^{1}$ and temporal $t\in T^{1}$ variables $q^{a}
(q^{1},\, q^{0})\equiv (q,\, t)\,\, (a=1,0),$ and to the formula for
interval $d\hat{\eta}^{2}=ds^{2}_{q}=dt^{2}-dq^{2},$ while the
ingredient spaces $ R_{(\pm)}^{1}$ are spanned by the coordinates
$\eta^{(\pm1)},$ respectively. Translating this into the language of
geometry of the Sect.3, upon reduction we may write
\begin{equation}
\begin{array}{l}
d\hat{\eta}=(O_{+}d\eta^{(+1)}+ O_{-}d\eta^{(-1)})\times
\sigma_{1}=\\d\hat{q}\equiv e_{0}dt + e_{q} dq, \label{RE1}
\end{array}
\end{equation}
where $e_{0}=\xi_{0}\times \sigma_{1}$ and $e_{q}=\xi\times
\sigma_{1}$ are, respectively, the temporal and spatial basis
vectors along the axes of $S_{(2)},$ and that $ q=
(1/\sqrt{2})(\eta^{(+1)}-\eta^{(-1)}),\quad t=
(1/\sqrt{2})(\eta^{(+1)}+\eta^{(-1)}),
 $
and  $v^{(\pm 1)}=(d\eta^{(\pm 1)}/dt)=(1/\sqrt{2})(1\pm v_{q}),
\,v_{q}=(dq/dt)=const.$ The law of inertia states that a free
particle in motion of uniform speed ($v_{q}=const$) in a straight
line in free space $R^{1}$ tends to stay in this motion and a
particle at rest tends to stay at rest unless acted upon by an
unbalanced force.  Below, we introduce the most important for the
treatment of inertia new concepts of {\em relative} state and
universal, so-called, {\em absolute}  state of ingredient space
$R_{(\pm)}^{1}$. The key measure for these states will be the
magnitude of the velocity components $(v^{(+1)},\,v^{(-1)})$ of
particle of interest.

{\em Definition.} The space $R_{(\pm)}^{1}$ is in the {\em absolute}
(abs) state if $v^{(\pm1)}=0;$  the space $R_{(\pm)}^{1}$ is in the
{\em relative} (rel) state if $v^{(\pm1)}\neq 0.$

According to it, the space $M_{2}$ can be realized respectively in
the following three states:  the {\em semi-absolute} states (rel,
abs) or (abs, rel), and {\em total relative } state (rel, rel). It
is remarkable that the {\em total-absolute} state (abs, abs) of
$M_{2}$, which is equivalent to the unobservable Newtonian {\em
absolute} two-dimensional spacetime, cannot be realized because of
$v^{+1}+v^{-1}=\sqrt{2}c$ (we re-instate the factor (c)), where
$v^{\pm1}\geq 0.$  The existence of {\em absolute} state of
$R_{(+)}^{1}$ is the immediate cause of the light traveling in free
space $R^{1}$ along $q$-axis with the resulting maximal velocity
$v_{q}=c,$ respectively, in $(+)$-direction in case of $
(v^{(+1)},\,0) \Leftrightarrow $ (rel, abs) and in $(-)$- direction
in case of $(0,\,v^{(-1)}) \Leftrightarrow$ {(abs, rel)}. Also note
that the {\em absolute} state of $R_{(+)}^{1}$ manifests its {\em
absolute} character in the important for special relativity fact
that the resulting velocity of light in free space $R^{1}$ is the
same for all inertial frames, that is, if $v^{(\pm1)}=0$ then
$v^{(\pm1)}=v^{(\pm1)}{}'=v^{(\pm1)}{}''=...=0.$ The velocity
$v^{(\pm1)}\neq 0$ is the measure of difference from the {\em
absolute} state. We might expect that this has a substantial effect
in alteration of particle motion under the unbalanced force. Similar
reasoning prompts us, further, to introduce the {\em distortion
inertial field} potential which depends on the rate of change of
this measure and allows the physical interpretation of the RLI as
follows:

RLI: {\em The rate of change of constant velocity $(v^{(\pm 1)})$
(both magnitude and direction) of massive $(m)$ test particle under
the unbalanced net force $(f)$ is the immediate cause of distortion
of local internal properties of flat space $M_{2} \rightarrow R_{2}$
conducted under the distortion inertial field potential}
\begin{equation}
\begin{array}{l}
{\ae}a_{(in)}^{(\pm1)}(q,t)=\pm \varrho(q,t,m,f) v^{(\pm1)}.
\end{array}
\label{RL00}
\end{equation}
The function $\varrho(q,t,m,f),$ which dependents on the rate of
change of the $v^{(\pm1)}$, will be determined below. Following
general prescription of Eq.~(\ref{R35}), the distortion
transformations of basis vectors $O_{\lambda}$ may be recast as
\begin{equation}
\begin{array}{l}
\widetilde{O}_{(+1)}={\cal Q}^{\tau}_{(+1)}(a)\,
O_{\tau}=O_{+}+{\ae}a_{(in)}^{(+1)}O_{-},\\
\widetilde{O}_{(-1)}={\cal Q}^{\tau}_{(-1)}(a)\,
O_{\tau}=O_{-}+{\ae}a_{(in)}^{(-1)}O_{+}. \label{RE55}
\end{array}
\end{equation}
Now let a second observer, who makes measurements using a frame of
reference $\widetilde{S}_{(2)}$ which is held stationary in
distorted space $R_{2}$, uses for the test particle the
real-curvilinear coordinates $\widetilde{q}{}^{a}(\widetilde{q},\,
\widetilde{t})$, where $\widetilde{q}=
(1/\sqrt{2})(\widetilde{\eta}{}^{(+1)}-\widetilde{\eta}{}^{(-1)}),\quad
\widetilde{t}=
(1/\sqrt{2})(\widetilde{\eta}{}^{(+1)}+\widetilde{\eta}{}^{(-1)}).$
The choice of ~(\ref{RE55}) has completely fixed the original form
of the interval we are to use $d\widetilde{s}{}^{2}_{q}\equiv
d\hat{\widetilde{\eta}}{}^{2}$, provided
\begin{equation}
\begin{array}{l}
d\widehat{\widetilde{\eta}}=(\widetilde{O}_{(+1)}d\widetilde{\eta}{}^{(+1)}+
\widetilde{O}_{(-1)}d\widetilde{\eta}{}^{(-1)})\times \sigma_{1}=\\
d\hat{\widetilde{q}}=\widetilde{e}_{0}d\widetilde{t} +
\widetilde{e}_{q} d\widetilde{q}, \label{RE4}
\end{array}
\end{equation}
where $\widetilde{e}_{0}=\widetilde{\xi}_{01}\times \sigma_{1}$ and
$\widetilde{e}_{q}=\widetilde{\xi}_{1}\times \sigma_{1}$ are,
respectively, the temporal and spatial basis vectors, and that
$\widetilde{\xi}_{01}=
(1/\sqrt{2})(\widetilde{O}_{(+1)}+\widetilde{O}_{(-1)}),\quad
\widetilde{\xi}_{1}=
(1/\sqrt{2})(\widetilde{O}_{(+1)}-\widetilde{O}_{(-1)}).$ The
Eq.~(\ref{RE3}) now gives
\begin{equation}
\begin{array}{l}
d\widetilde{\eta}^{A}=\frac{\partial \widetilde{\eta}^{A}}{\partial
\eta^{C}}d\eta^{C}=\\\left[<\widetilde{O}^{A},O_{C}>
+<\widetilde{O}^{A},\chi_{C}(e,F)>\right]d\eta^{C},
\end{array}
\label{RL3}
\end{equation}
where the capital Latin indices $A,C,$ etc. run over $(\pm 1).$ In
terms of corresponding matrices $(\cdots)$ it can be re-written as
\begin{equation}
(d\widetilde{\eta}^{A})=\left[(<\widetilde{O}^{A},O_{C}>)
+(<\widetilde{O}^{A},\chi_{C}>)\right](d\eta^{C}), \label{RL4}
\end{equation}
where $(d\widetilde{\eta}^{A})= \left(
  \begin{array}{c}
    d\widetilde{\eta}{}^{(+1)} \\
    d\widetilde{\eta}{}^{(-1)} \\
  \end{array}
\right) $ and so on. \, Denoting $\, {\cal
Q}_{\eta}=(<\widetilde{O}_{A},O^{C}>),$ we obtain
\begin{equation}
\begin{array}{l}
(<\widetilde{O}^{A},O_{C}>)={\cal Q}^{-1}_{\eta}=\\\gamma_{a} \left(
  \begin{array}{cc}
    1 & -{\ae}a_{(in)}^{(+1)} \\
    -{\ae}a_{(in)}^{(-1)} & 1 \\
  \end{array}
\right), \label{RL5}
\end{array}
\end{equation}
where $\gamma_{a}= (1-{\ae}^{2}a_{(in)}^{(+1)}
a_{(in)}^{(-1)})^{-1}=(1+\varrho^{2}/2\gamma^{2}_{q})^{-1},$ $
\gamma_{q}=(1-v_{q}^{2})^{-1/2}.$ The transformation equation for
coordinates becomes
\begin{equation}
\begin{array}{l}
d\widetilde{\eta}{}^{(\pm 1)}=\gamma_{a} (d\eta^{(\pm 1)} \mp
\varrho v^{(\pm1)}d\eta^{(\mp 1)}),
\end{array}
\label{RL7}
\end{equation}
while $\chi_{(\pm 1)}=e_{(\pm 1)}\omega(0)=0.$ Actually, the
temporal $ {\ae}a_{(in)}^{0}= \varrho v_{q}$ and spatial
${\ae}a_{(in)}^{1}= \varrho$ components of inertial field yield
$F_{10}=\partial_{q}(\varrho v_{q})-\partial_{0}\varrho=0.$ Then,
$d\hat{\widetilde{q}}=d\hat{q}$ or
$d\widetilde{s}{}^{2}_{q}=ds^{2}_{q}=dt^{2}/\gamma^{2}_{q}$, that
is, $g_{ab}=(\partial q^{c}/\partial\widetilde{q}^{a})(\partial
q^{d}/\partial\widetilde{q}^{b})\eta_{cd},\,$ ($a,b,c,d=0,1$), where
the components of the metrical tensor $\eta_{cd}$ assume the values
$(-1,1, 0).$ Introducing new coordinates
$d\widetilde{\eta}'{}^{A}=\gamma_{a}^{-1} d\widetilde{\eta}{}^{A},$
the metrical tensor transformed to $g'_{AB}=(\partial
\widetilde{\eta}{}^{C}/\partial \widetilde{\eta}'{}^{A}) (\partial
\widetilde{\eta}{}^{D}/\partial
\widetilde{\eta}'{}^{B})g_{CD}=\gamma_{a}^{2}g_{AB},$ where
$g_{AB}=D_{A}^{C}D_{B}^{C}.$ To retain former notational
conventions, from now on we will omit the prime at the quantities
$\widetilde{\eta}'{}^{(A)},$ $\widetilde{q}',$ $g'_{AB},...$ The
Eq.~(\ref{RL7}) then becomes
\begin{equation}
\begin{array}{l}
d\widetilde{\eta}{}^{(\pm 1)}=d\eta^{(\pm 1)} \mp \varrho
v^{(\pm1)}d\eta^{(\mp 1)}= (v^{(\pm1)}\mp
\frac{\varrho}{2\gamma^{2}_{q}})dt.
\end{array}
\label{RL777}
\end{equation}
The generally covariant expression for interval is
$d\widetilde{s}{}^{2}_{q}=(d\hat{\widetilde{q}})^{2}=g_{ab}d\widetilde{q}{}^{a}d\widetilde{q}{}^{b},$
provided
\begin{equation}
\begin{array}{l}
g_{00}=(1+\frac{\varrho^{2}}{2\gamma^{2}_{q}})^{-2}[(1+\frac{\varrho v_{q}}{\sqrt{2}})^{2}+\frac{\varrho^{2}}{2}],\\
g_{11}=-(1+\frac{\varrho^{2}}{2\gamma^{2}_{q}})^{-2}[(1-\frac{\varrho
v_{q}}{\sqrt{2}})^{2}-\frac{\varrho^{2}}{2}],\\
g_{10}=g_{01}=-\sqrt{2}\varrho\,(1+\frac{\varrho^{2}}{2\gamma^{2}_{q}})^{-2}.
\label{RE494}
\end{array}
\end{equation}
It is easily verified that the resulting curvature is not zero
 because of
inequalities $\partial_{q}\psi^{0}_{0}\neq
\Gamma^{0}_{01}\psi^{0}_{0},$ or $\partial_{q}\psi^{1}_{1}\neq
\Gamma^{1}_{11}\psi^{1}_{1},$ etc, where
$\psi^{\mu}_{l}=D^{\mu}_{l}$. The difference of the vectors
$d\widehat{\eta}$ ~(\ref{RE1}) and
$d\widehat{\widetilde{\eta}}$~(\ref{RE4}) could be interpreted by
the second observer as being due to the distortion of basis vectors
$O_{\pm}$ of flat space. However, this difference with equal justice
could be interpreted by him as a definite criterion for the {\em
absolute} character of his own state of acceleration, rather than to
any absolute quality of distortion of local internal properties of
flat space. To prove this assertion, we may derive from the
Eq.~(\ref{RL777}) the general transformation equations for spatial
and temporal intervals to accelerated axes as
\begin{equation}
\begin{array}{l}
d\widetilde{q}= dq(1+\frac{\varrho v_{q}
}{\sqrt{2}})-\frac{\varrho }{\sqrt{2}}dt,\\
d\widetilde{t}=dt(1-\frac{\varrho v_{q}}{\sqrt{2}})+\frac{\varrho
}{\sqrt{2}}dq.
\end{array}
\label{RL8}
\end{equation}
The foregoing transformation equations give a reasonable change at
low velocities $\varrho /\sqrt{2} <<1 \quad (v_{q}\sim 0):\,$
$d\widetilde{q}\simeq dq-(\varrho /\sqrt{2})dt,\quad
d\widetilde{t}\simeq dt,$ which become conventional transformation
equations to accelerated $(a\neq 0)$ axes at
$\varrho/\sqrt{2}=\int_{0}^{t} a dt'$. This immediately indicates
that we may introduce (in Newton's terminology)  the {\em absolute}
acceleration as
\begin{equation}
\begin{array}{l}
\textbf{a}_{abs}\equiv e_{q}\frac{d\varrho }{\sqrt{2}ds_{q}}.
\end{array}
\label{RL888}
\end{equation}
We may also introduce the, so-called, {\em inertial} acceleration
\begin{equation}
\begin{array}{l}
\textbf{a}_{in}\equiv e_{q}a^{1}=e_{q}
\frac{d^{2}\widetilde{q}}{ds^{2}_{q}}=e_{q}
\frac{1}{\sqrt{2}}(\frac{d^{2}\widetilde{\eta}{}^{(+1)}}{ds^{2}_{q}}-
\frac{d^{2}\widetilde{\eta}{}^{(-1)}}{ds^{2}_{q}}).
\end{array}
\label{RL88}
\end{equation}
Combining  ~(\ref{RL777}), (\ref{RL888}) and ~(\ref{RL88}), we
obtain the key relation
\begin{equation}
\begin{array}{l}
\gamma_{q}\textbf{a}_{in}=-\textbf{a}_{abs}.
\end{array}
\label{RL999}
\end{equation}
Suppose the position of test particle in the space $M_{4}$, in
general, is specified by the coordinates $x^{l}(s)$ $ (l=1,2,3,0)$
with respect to the axes of inertial system $S_{(4)}.$ The specific
problem that now arises is to obtain equations connecting the {\em
absolute} acceleration (Eq.~(\ref{RL888})) given in the inertial
system $S_{(2)}$ to the unbalanced relativistic force~(\cite{W72}):
$ f^{l}(\textbf{f},\,f^{0})=m(d^{2}x^{l}/ds^{2})=
\Lambda^{l}_{k}(\textbf{v})F^{k}, $ exerted on the test particle.
Here $F^{k}(\textbf{F},0)$ is the force in the proper reference
frame of test particle, $\Lambda^{l}_{k}(\textbf{v})$ is the Lorentz
transformation matrix $(i,j=1,2,3)$: $ \Lambda^{i}_{j}=\delta_{i
j}-v_{i}v_{j}(\gamma-1)/\textbf{v}^{2},$ and $
\Lambda^{0}_{i}=\gamma v_{i},$ where $
\gamma=(1-\textbf{v}^{2})^{-1/2}.$ Then the two systems can be
chosen so that the axis $e_{q}$ of $S_{(2)}$ lies along the acting
force $\textbf{f}=e_{f}|\textbf{f}|$ while the time coordinates in
the two systems are taken the same $q^{0}=x^{0}=t.$ This choice
($e_{q} = e_{f}$) of unit vectors, which can always be made, implies
$v_{q}=(e_{f}\cdot \textbf{v})$, and that the rate of change of the
measure of difference from the {\em absolute} state of massive ($m$)
test particle under the unbalanced net force $f^{l}(\textbf{f},
f^{0})$ other than gravitational can be determined as
\begin{equation}
\begin{array}{l}
\frac{1}{\sqrt{2}}\frac{d\varrho}{ds_{q}}=\frac{1}{m}|f^{l}|=\frac{1}{m\gamma}
|\textbf{f}|.
\end{array}
\label{RLL9}
\end{equation}
The key relation ~(\ref{RL999}) provides quantitative means for the
RLI as
\begin{equation}
\begin{array}{l}
\textbf{f}_{(in)}=m\textbf{a}_{in}=
e_{q}(-m\Gamma^{1}_{ab}(\varrho)\frac{d\widetilde{q}{}^{a}}{d\widetilde{s}_{q}}
\frac{d\widetilde{q}^{b}}{d\widetilde{s}_{q}})=\\
-m\textbf{a}_{abs}/\gamma_{q}=-[\textbf{F}+(\gamma-1)\textbf{v}(\textbf{v}\cdot\textbf{F})/v^{2}]/\gamma_{q}\gamma,
\end{array}
\label{RLL999}
\end{equation}
where $\textbf{f}_{(in)}$ is the {\em inertial} force, also we have
taken into account that the trajectory of the particle in curved
space $R_{2}$ is given by the equation for the geodesic. At low
velocities $v_{q}\simeq |\textbf{v}|\simeq 0,$  the
Eq.~(\ref{RLL999}) reduces to conventional law of inertia
\begin{equation}
\begin{array}{l}
\textbf{f}_{(in)}=-\textbf{F}.
\end{array}
\label{RLG7}
\end{equation}
At high velocities $|\textbf{v}|\sim 1,$ we have $e_{f}\simeq
e_{v},$ where $\textbf{v}=e_{v}|\textbf{v}|,$ and that
$v_{q}\simeq|\textbf{v}|\sim 1.$ The Eq.~(\ref{RLL999}) then gives
\begin{equation}
\begin{array}{l}
\textbf{f}_{(in)}\simeq-e_{v}(e_{v}\cdot\textbf{F})/\gamma,
\end{array}
\label{RKK}
\end{equation}
which vanishes in the limit of the photon $(m=0).$ This can be
easily understood. Certainly, the acceleration of photon in free
space along its traveling direction is impossible, as well as its
deflection by some force is also discarded  because otherwise its
velocity becomes greater than $(c).$ Thus, it takes force to disturb
inertia state, i.e. to make {\em absolute} ($\textbf{a}_{abs}$)
acceleration. The {\em absolute} acceleration is due to the
existence of the {\em absolute} state of ingredient space
$R_{(\pm)}^{1}$ and, evidently, it is admitted as {\em the immediate
cause of the real distortion of the local internal properties of
flat space $M_{2}$.} The {\em relative} ($d\varrho/ds_{q}=0$)
acceleration (in Newton's terminology) (both magnitude and
direction), in contrary, cannot be the cause of the distortion of
the space and, thus, it does not produce inertia effect.

\subsection{Involving gravitation; the Principle of Equivalence}
In the development of the relativistic field theory of inertia in
the more general case when gravitational action is involved, we are
at once led to seek equations in the form of generally covariant
tensor expression using any set of general coordinates which we may
desire to introduce. Let the distortion gauge field $(a_{l})$
underlies gravitation. Then,  the generally covariant expression for
interval in four-dimensional Riemannian space $R_{4}$ is
$d\widetilde{s}^{2}=g_{\mu\nu}(a)d\widetilde{x}{}^{\mu}
d\widetilde{x}{}^{\nu},$  and $\Gamma^{\lambda}_{\mu\nu}(a)$ denotes
affine connection agreed with the metric $g_{\mu\nu}(a)$. Although
inertial forces do not exactly cancel for freely falling systems in
an inhomogeneous  or time-dependent gravitational field, in
accordance with the Principle of Equivalence, we may still maintain
for a sufficiently small region that the effects of gravitation
could be removed by the use of local freely falling coordinate frame
$S_{4}^{(l)}$, having the natural acceleration due to gravity for
that region. This implies that the local spacetime structure can be
identified with the Minkowski spacetime possessing Lorentz symmetry,
and that the physical laws in the frame $S_{4}^{(l)}$ will be the
same as in any inertial coordinate frame $S_{(4)}$. This is similar
to assumption of approximate replacement of a curved surface by its
tangent plane at a given point, made use of in geometrical
considerations. Therefore, we can always choose {\em natural
coordinates} $X^{\alpha}(X,Y,Z,T)=(\textbf{X},\,T)$ with respect to
the axes of the system $S_{4}^{(l)}$ in immediate neighbourhood of
any spacetime point $(\widetilde{x}_{p})\in R_{4}$ in question, over
a differential region taken small enough so that we can neglect the
spatial and temporal variations of gravity for the range involved.
In this coordinates, the special relativity formula for interval
will be $dS^{2}=\eta_{lk}dX^{l}dX^{k}=d\widetilde{s}^{2},$ where the
components of the metrical tensor $\eta_{lk}$ assume the special
relativity values $(-1,1, 0),$ and the first differential
coefficients of the $\eta_{lk}$ with respect to these coordinates
will be zero at the point $(p).$ In general, however, the second
differential coefficients will not be zero except for the special
case of spacetime that actually is flat. The values of metrical
tensor $g_{\mu\nu}(a)$ and affine connection
$\Gamma^{\lambda}_{\mu\nu}(a)$ at the point $(\widetilde{x}_{p})$ is
necessarily sufficient information for determination of the natural
coordinates $X^{\alpha}(\widetilde{x}{}^{\mu})$ in the small region
of neighbourhood of selected point~\citep{W72}. The whole scheme
outlined in the previous subsection will then hold in the frame
$S_{4}^{(l)}$, provided as a preliminary step we first examine the
possibility of re-expressing the special relativity formula for the
free $(\textbf{f}_{(l)}=0)$ test particle $dU^{\alpha}/dS$
$=d^{2}X^{\alpha}/dS^{2}=0,$ $(\alpha=1,2,3,0)$ in generally
covariant form as $DU^{\alpha}/D\widetilde{s}$
$=D^{2}X^{\alpha}/D\widetilde{s}^{2}=0,$ where $\textbf{f}_{(l)}$ is
the special relativity value of unbalanced relativistic force other
than gravitational in the frame $S_{4}^{(l)}$ and, according to the
general prescription $D/D\widetilde{s}$ is the covariant derivative
along the curve $\widetilde{x}^{\mu}(\widetilde{s})\in R_{4}.$ The
relativistic gravitational force $f^{\mu}_{g}(\widetilde{x})$
exerted on the test particle of the mass $(m)$ is
\begin{equation}
\begin{array}{l}
f^{\mu}_{g}(\widetilde{x})=m\frac{d^{2}\widetilde{x}{}^{\mu}}{d\widetilde{s}^{2}}=-
m\Gamma^{\mu}_{\nu\lambda}(a)\frac{d\widetilde{x}{}^{\nu}}{d\widetilde{s}}
\frac{d\widetilde{x}^{\lambda}}{d\widetilde{s}}.
\end{array}
\label{RG1}
\end{equation}
The frame $S_{4}^{(l)}$ will be valid if only the gravitational
force given in this coordinate frame
\begin{equation}
\begin{array}{l}
f^{\alpha}_{g(l)}=\frac{\partial X^{\alpha}}{\partial
\widetilde{x}{}^{\mu}}f^{\mu}_{g},
\end{array}
\label{RG15}
\end{equation}
could be removed by the inertial force which, in turn, is due to the
distortion inertial field potential Eq.~(\ref{RL00}). Whereas, the
two systems $S_{2}$ and $S_{4}^{(l)}$ can be chosen so that the axis
$e_{q}$ of $S_{(2)}$ now lies along ($e_{q} =e_{f}$) the acting net
force $\textbf{f}= \textbf{f}_{(l)}+\textbf{f}_{g(l)}(a),$ while the
time coordinates in the two systems are taken the same
$q^{0}=t=X^{0}=T.$ Instead of Eq.~(\ref{RLL9}), we now have
\begin{equation}
\begin{array}{l}
\frac{1}{\sqrt{2}}\frac{d\varrho}{ds_{q}}=
\frac{1}{m}|f^{\alpha}_{(l)}+f^{\alpha}_{g(l)}(a)|.
\end{array}
\label{RLL12G}
\end{equation}
Hence, in general,  the RLI can be obtained as
\begin{equation}
\begin{array}{l}
\textbf{f}_{(in)}=m\textbf{a}_{in}=
e_{q}(-m\Gamma^{1}_{ab}(\varrho)\frac{d\widetilde{q}{}^{a}}{d\widetilde{s}_{q}}
\frac{d\widetilde{q}^{b}}{d\widetilde{s}_{q}})=\\
-\frac{m\textbf{a}_{abs}}{\gamma_{q}}=
-\frac{e_{f}}{\gamma_{q}}|f^{\alpha}_{(l)}- m\frac{\partial
X^{\alpha}}{\partial
\widetilde{x}{}^{\sigma}}\Gamma^{\sigma}_{\mu\nu}(a)
\frac{d\widetilde{x}{}^{\mu}}{dS} \frac{d\widetilde{x}^{\nu}}{dS}|.
\end{array}
\label{RLLL9}
\end{equation}
In spite of totally different and independent sources of gravitation
and inertia,  at $\textbf{f}_{(l)}=0$ when the mass $(m)$ is
canceled out in Eq.(\ref{RLLL9}), the RLI  indeed furnishes
justification for introduction of the Principle of Equivalence. A
remarkable feature is that the inertial force is of the same nature
as gravitational force, i.e., both are due to the distortion of
local internal properties of the flat space. The non-vanishing
inertial force acting on the photon of energy $h\nu$, and that of
effective mass $\left(h\nu/c^{2}\right)$ after inserting a factor
$(c^{2})$ which so far was suppressed, can be obtained from the
Eq.~(\ref{RLLL9}), at $\textbf{f}_{(l)}=0,$ as
\begin{equation}
\begin{array}{l}
\textbf{f}_{(in)}= -\left(\frac{h\nu}{c^{2}}\right)e_{f}
|\frac{\partial X^{\alpha}}{\partial
\widetilde{x}{}^{\sigma}}\Gamma^{\sigma}_{\mu\nu}(a)
\frac{d\widetilde{x}{}^{\mu}}{dT}
\frac{d\widetilde{x}^{\nu}}{dT}|=\\
-\left(\frac{h\nu}{c^{2}}\right)e_{f}|(\frac{d^{2}\widetilde{t}}{dT^{2}})
\frac{dX^{\alpha}}{d\widetilde{t}}+(\frac{d\widetilde{t}}{dT})^{2}\frac{\partial
X^{\alpha}}{\partial \widetilde{x}^{i}}\frac{d
u_{i}}{d\widetilde{t}}|,
\end{array}
\label{RLLL10}
\end{equation}
provided $e_{f}=(\textbf{X}/|\textbf{X}|)$, $v_{q}=(e_{f}\cdot
\textbf{u})=|\textbf{u}|,\,\,(\gamma_{q}=\gamma)$ where
($\textbf{u}$) is the velocity of photon and
 $(d \textbf{u}/d\widetilde{t})$ is the
acceleration, and that, $g_{\mu\nu}(a)(d\widetilde{x}{}^{\mu}/dT)
(d\widetilde{x}^{\nu}/dT)=0.$ To obtain some feeling for this, we
may turn now to the (PPN)
approximation~\citep{will1,will2,Nord,will3}, which can be regarded
as a deformation of a background asymptotically flat Minkowski
metric. In this context we calculate the inertial force for the
photon in gravitating system of particles that are bound together by
their mutual gravitational attraction to order $\bar{v}^{2}\sim
G_{N}\bar{M}/\bar{r}$ of small parameter, where $\bar{v},\,\bar{M}$
and $\bar{r}$ are typical , respectively, the average values of
their velocities, masses and separations. In doing this, we may
expand the metrical tensor to the following order: $
g_{00}=1+\stackrel{2}{g}_{00}+\stackrel{4}{g}_{00}+...,\quad
g_{ij}=-\delta_{ij}+\stackrel{2}{g}_{ij}+\stackrel{4}{g}_{ij}+...,\quad
 g_{i0}=\stackrel{3}{g}_{i0}+\stackrel{5}{g}_{i0}+....,
$ where $\stackrel{N}{g}_{\mu\nu}$ denotes the term of order
$\bar{v}^{N}.$ Taking into account the standard  expansions of
affine connection
$\Gamma^{\sigma}_{\mu\nu}=\stackrel{2}{\Gamma^{\sigma}_{\mu\nu}}+\stackrel{4}{\Gamma^{\sigma}_{\mu\nu}}+...$
for the components $\Gamma^{i}_{00},\,\Gamma^{i}_{jk},\,
\Gamma^{0}_{0i},$ and that
$\Gamma^{\sigma}_{\mu\nu}=\stackrel{3}{\Gamma^{\sigma}_{\mu\nu}}+\stackrel{5}{\Gamma^{\sigma}_{\mu\nu}}+...$
for the components $\Gamma^{i}_{0j},\,\Gamma^{0}_{00},\,
\Gamma^{0}_{ij},$ where
$\stackrel{2}{\Gamma^{i}_{00}}=\stackrel{2}{\Gamma^{0}_{0i}}
=-(1/2)(\partial\stackrel{2}{g}_{00}/\partial \widetilde{x}{}^{i})$
etc., hence to the required accuracy we obtain
\begin{equation}
\begin{array}{l}
\stackrel{2}{\textbf{f}}_{(in)}=
-\left(\frac{h\nu}{c^{2}}\right)e_{f} |\stackrel{1}{(\frac{\partial
X^{\alpha}}{\partial \widetilde{x}{}^{\sigma}})}
\stackrel{2}{(\frac{d^{2}\widetilde{x}{}^{\sigma}}{dT^{2}})}|=
-\left(\frac{h\nu}{c^{2}}\right)
\stackrel{2}{(\frac{d\textbf{u}}{d\widetilde{t}})}=\\
-\left(\frac{h\nu}{c^{2}}\right)[-2\bf{\nabla}\phi+4\textbf{u}(\textbf{u}\cdot
\mathbf{\nabla}\phi) + O(\bar{v}{}^{3})],
\end{array}
\label{RG2}
\end{equation}
where $\phi$ is the Newton's potential, such that
$\stackrel{2}{g}_{00}=2\phi,$ $
\stackrel{2}{g}_{ij}=2\delta_{ij}\phi,$ and
$|\textbf{u}|=1+2\phi+O(\bar{v}{}^{3}).$

\section{The rearrangement of vacuum state}
Collecting together the results just established in previous
sections we finally arrive at discussion of the rearrangement of
vacuum in gravitation. To trace this line,  in realization of the
gravitation gauge group $G_{R}$ we implement the  abelian local
group
\begin{equation}
\begin{array}{l}
U^{loc}=U(1)_{Y}\times \overline{U}(1)\equiv U(1)_{Y}\times
diag[SU(2)],
\end{array}
\label{R45}
\end{equation}
with the group elements of $exp\,[i\frac{Y}{2}\,\theta_{Y}(\eta)]$
of $U(1)_{Y}$ and $exp\,[iT^{3}\,\theta_{3}(\eta)]$ of
$\overline{U}(1)$. The group Eq.~(\ref{R45}) leads to the
renormalizable theory on $M_{6}$ because gauge invariance gives
conservation of charge, also ensures the cancelation of quantum
corrections that would otherwise result in infinitely large
amplitudes. This has two generators, the third component $T^{3}$ of
isospin $\mathbf{T}$ related to the Pauli spin matrix
$\frac{\mathbf{\tau}}{2}$, and hypercharge $Y$ implying
$$
\begin{array}{l}
Q^{d}=T^{3}+\frac{Y}{2},
\end{array}
$$ where $Q^{d}$ is the {\em distortion
charge} operator assigning the number -1 to particles, but +1 to
anti-particles. The group Eq.~(\ref{R45}) entails two neutral gauge
bosons $W^{3}_{A}$ of $\overline{U}(1)$ or that coupled to $T^{3}$,
and $B_{A}$ of $U(1)_{Y}$, or that coupled to hypercharge $Y$. Gauge
invariant Lagrangian of fermion field is given in standard form $
{\cal L}=\overline{\psi}(\eta)i\gamma^{A}\,D_{A}\psi(\eta), $
provided by covariant derivative $
D_{A}\,\psi(\eta)=\left(\partial_{A}-ig\,T^{3}\,W^{3}_{A}-
ig'\,(Y/2)\,B_{A}\right)\,\psi(\eta), $ and $g,\,g'$ being the
$\overline{U}(1)$,$\quad U(1)_{Y}$ coupling strength, respectively.
Spontaneous symmetry breaking can be achieved by introducing the
neutral complex scalar Higgs field
$$
\begin{array}{l}
\phi= \left(
  \begin{array}{c}
    0 \\
    \phi^{0} \\
  \end{array}
\right) , \quad Y(\phi)=1, \quad
\phi^{0}=\frac{1}{\sqrt{2}}(\phi_{1}+i\phi_{2}),
\end{array}
$$ with the standard
potential energy density function $ V(\phi)=-\mu^{2}\phi^{+}\,\phi
+\lambda \left(\phi^{+}\,\phi\right)^{2}, $ where $\mu^{2}>0,\quad
\lambda>0$. This is ingredient of the gauge invariant Lagrangian of
Higgs field $ {\cal L}_{H}=
\left(D_{A}\,\phi\right)^{+}\left(D^{A}\,\phi\right)-V(\phi), $
where $D_{A}\,\phi(\eta)=(\partial_{A}-ig\,T^{3}\,W^{3}_{A} -
ig'\,(Y/2)\,B_{A})\,\phi(\eta).$ Minimization of the vacuum energy
fixes non-vanishing  VEV:
$$
\begin{array}{l}
<\phi>_{0}\equiv<0|\phi|0>= \left(
  \begin{array}{c}
    0 \\
    \frac{v}{\sqrt{2}} \\
  \end{array}
\right) , \quad v=\left(\frac{\mu^{2}}{\lambda}\right)^{1/2},
\end{array}
$$
 leaving one
Goldstone boson. The VEV of spontaneously breaks the theory, leaving
the $U(1)_{d}$ subgroup intact. The unitary gauge
$$
\begin{array}{l}
\phi(\eta)=U^{-1}(\xi_{3}) \left(
  \begin{array}{c}
    0 \\
    \frac{v+\zeta(\eta)}{\sqrt{2}} \\
  \end{array}
\right), \, U(\xi_{3})=\exp \left[ \frac{i\xi_{3}\cdot
\tau^{3}}{v}\right],
\end{array}
$$
is parameterized by two real shifted fields $\xi_{3}$ and $\zeta$,
such that $<0|\xi_{3}|0>=<0|\zeta|0>=0$. The gauge transformation
$$
\begin{array}{l}
\phi'=U(\xi_{3})\,\phi=\frac{v+\zeta}{\sqrt{2}}\,\chi, \quad \chi=
\left(
  \begin{array}{c}
    0 \\
    1 \\
  \end{array}
\right),
\end{array}
$$ leads to
$ V(\phi')=\mu^{2}\zeta^{2}+\lambda v
\zeta^{3}+(\lambda/4)\,\zeta^{4}, $ which gives the mass of Higgs
boson $M_{H}=\sqrt{2}\,\mu.$ An examination of the $v^{2}$-terms in
the kinetic piece of the Lagrangian $ {\cal L}_{H}=
\left(D_{A}\,\phi'\right)^{+}\left(D^{A}\,\phi'\right)-V(\phi')$
 reveals the mass terms for the
physical gauge bosons:
\begin{equation}
\begin{array}{l}
\frac{v^{2}}{2}\,\left|\left(i\frac{g}{2}\,\tau^{3}\,W'{}^{3}_{A}+
ig'\,\frac{Y}{2}\,B_{A}'\right)\,\chi\right|^{2}=\\
\frac{1}{2}\,\left(Z_{A}, \,A_{A}\right) \left(
  \begin{array}{cc}
    M^{2}_{Z}  & 0 \\
    0 & 0 \\
  \end{array}
\right) \left(
  \begin{array}{c}
    Z^{A}  \\
    A^{A} \\
  \end{array}
\right).
\end{array}
\label{R52}
\end{equation}
The mass matrix can be diagonalized by the standard orthogonal
transformations:
\begin{equation}
\begin{array}{l}
Z_{A}= \cos \theta_{W}\,W'{}^{3}_{A}+\sin \theta_{W}\,B'_{A},\\
A_{A}= \sin \theta_{W}\,W'{}^{3}_{A}+\cos \theta_{W}\,B'_{A},\\
M_{Z}=\frac{v}{2}\sqrt{g^{2}+g'{}^{2}}, \quad M_{A}=0,
\end{array}
\label{R53}
\end{equation}
where $\tan \theta_{W}=g'/g$. Namely, the neutral gauge field
$W'{}^{3}_{A}$ mixes with the  abelian gauge field $B_{A}'$ to form
the physical states $Z_{A}$ and $A_{A}$, with the masses $M_{Z}$ and
$M_{A}$, respectively. For neutral current we get
\begin{equation}
\begin{array}{l}
{\cal L}_{int}=\ae\,\left({\cal J}^{(0)}_{A}\,A^{A}+{\cal
J}^{(M)}_{A}\,Z^{A}\right)\equiv \ae\,\left({\cal
J}_{A}\,a^{A}\right) ,
\end{array}
\label{R54}
\end{equation}
where $\ae = g\,\sin \theta$, and
\begin{equation}
\begin{array}{l}
{\cal J}^{(0)}_{A}=\overline{\psi}(\eta)i
\gamma_{A}\,Q^{d}\,\psi(\eta),\\ {\cal
J}^{(M)}_{A}=\overline{\psi}(\eta)i
\gamma_{A}\,Q^{(in)}\,\psi(\eta),\\
{\cal J}_{A}=\overline{\psi}(\eta)i \gamma_{A}\,\psi(\eta), \quad
Q^{(in)}=\frac{T^{3}-\sin^{2} \theta_{W},\, Q^{gr}}{\sin
\theta_{W}\,\cos \theta_{W}}, .
\end{array}
\label{R55}
\end{equation}
These relations show that an additional substantial change of
properties of the spacetime continuum besides the curvature may be
arisen at huge energies. A more detailed analysis and calculations
on this will be presented on the later date.

\section{Concluding remarks}
Overall, we would sum up our investigation as follows. Following the
powerful method of phenomenological Lagrangians, in the framework of
GGP we connect the gravitation gauge group $G_{R}$ to nonlinear
realization of the Lie group $G_{D}$ of distortion of the flat space
$M_{6}$. The fundamental fields are distortion gauge fields, and
that the metric and connection are related to these gauge fields.
The agreement between this theory and observation is satisfactory.
On the suggested theoretical basis, we construct the relativistic
field theory  of inertia which treats inertia as a distortion of
local internal properties of flat space $M_{2}$. We derive the RLI
which furnishes justification for introduction of Principle of
Equivalence. Finally, we address the rearrangement of vacuum state
in gravity. Whereas, the missing ingredient is the Higgs boson. The
principle assumption went into the building of this approach that
the Higgs boson is coupled only with distortion field, but not with
the matter fields. The matter fields have interacted with the Higgs
boson only via the metrical tensor. The four parameters
$g,\,g',\,\mu,\,\lambda$ are inserted by hand, which consequently
determine two coupling constants $\ae_{A}$, $\ae_{Z}$ , and two
masses $M_{H}$, $M_{Z}$. However, two relations can be imposed upon
these parameters. First, the Compton wave-length $\lambda_{Z}$ of
massive component $Z_{a}$ is finite, which will be of vital interest
for the physics of superdense matter in very compact astrophysical
sources if, for example, we set $
\lambda_{Z}=2\hbar/cv\sqrt{g^{2}+g'{}^{2}}\leq 0.4 \, fm, $ where
$\simeq 0.4 \,fm$ is the distance between the particles at nuclear
density (we re-instate $\hbar$ and $c$). Second, we have $
2\sqrt{\pi\,G_{N}}=g\,g'/\sqrt{g^{2}+g'{}^{2}}.$

\acknowledgments
A knowledgable comments and useful suggestions from
the anonymous referee are much appreciated.

\appendix

\section{Further topics on the GGP }
\subsection{Field equations}
Field equations can be derived from an invariant action
\begin{equation}
\begin{array}{l}
S=S_{a} + S_{\widetilde{\Phi}}=\int \sqrt{-\eta}\, L_{a}\,d^{4}x +
\int \sqrt{-g}\, L_{\widetilde{\Phi}} \,d^{4}\widetilde{x},
\end{array}
\label{R15}
\end{equation}
where $L_{a}$ is the Lagrangian of distortion field ($a$),
$L_{\widetilde{\Phi}}$ is the Lagrangian of matter fields, whereas
the dependence on distortion gauge field comes only through the
components of metrical tensor. The $L_{a}$ is invariant under
($\Lambda$)-coordinate and $U^{loc}$-gauge groups. The Lagrangian
$L_{\widetilde{\Phi}}$, in turn, is invariant under gauge group of
gravitation $G_{R}$. In terms of Euler-Lagrange variations in
$M_{4}$ and $R_{4}$, we readily obtain
\begin{equation}
\begin{array}{l}
\frac{\delta \left(\sqrt{-\eta}\,L_{a}\right)}{\delta\,a_{l}}=-
\frac{\partial\, g^{\mu\nu}}{\partial\, a_{l}} \frac{\delta
(\sqrt{-g}\, L_{\widetilde{\Phi}})}{\delta\, g^{\mu\nu}}=-
\frac{\sqrt{-g}}{2} \, \frac{\partial \,g^{\mu\nu}}{\partial\,
a_{l}}\widetilde{T}_{\mu\nu},\\
\frac{\delta
\widetilde{L}_{\widetilde{\Phi}}}{\delta\,\widetilde{\Phi}}=0, \quad
\frac{\delta \,L_{\widetilde{\Phi}}}{\delta
\overline{\widetilde{\Phi}}}=0,
\end{array}
\label{R16}
\end{equation}
where $\widetilde{T}_{\mu\nu}$ is the energy-momentum tensor of
matter fields $\widetilde{\Phi}(\widetilde{x})$.

\subsection{The unitary map matrix}
In this subsection we will determine the unitary map matrix $R(a)$
and gauge invariant scalar function $S(F)$ for the fields of spin
$0,\,1,$ and $1/2$. Our strategy is as follows: we, first, insert
Eq.~(\ref{R3}) into Eq.~(\ref{R4}) to obtain an identity, and then
equate the coefficients in front of $\partial \,\Phi$ and $\Phi$ to
zero. In this way we may obtain the required relations to determine
$R(a)$ and $S(F)$.

1) A straightforward calculation for the fields of spin $j=0,1$
gives the unitary  matrix
\begin{equation}
\begin{array}{l}
R(\widetilde{x},x)=R(x)\,R_{g}(\widetilde{x})=\exp\left(
-i\Theta(x)- \Theta_{g}(\widetilde{x})\right), \end{array}
\label{R21}
\end{equation}
provided
\begin{equation}
\begin{array}{l}
\Theta(x)= \ae\int_{0}^{x}\,a_{l}(x)\,d\,x^{l},\\
\Theta_{g}(\widetilde{x})=  \int_{0}^{\widetilde{x}}\left[ R^{+}
\,\Gamma_{\mu} \,R+ \psi^{-1}\widetilde{\partial}_{\mu}
\,\psi\right]\,d\,\widetilde{x}{}^{\mu},
\end{array}
\label{R22}
\end{equation}
where $\psi \equiv \left(\psi^{\mu}_{l}\right),\quad \Theta_{g}=0\,$
for scalar field, and $\Theta_{g}+\Theta_{g}^{+}=0\,$ for vector
field because of $\Gamma_{\mu}+\Gamma_{\mu}^{+}=0$, $\Gamma_{\mu}$
is the connection. The scalar function $S(F)$ is given in the
Eq.~(\ref{R9}):
\begin{equation}
\begin{array}{l}
S(F)=\frac{1}{4}\,R^{+}(\psi^{l}_{\mu}\,D_{l}^{\mu})\,R =
\frac{1}{4}\,\psi^{l}_{\mu}\,D_{l}^{\mu}.
\end{array}
\label{RG9}
\end{equation}

2) Unitary map of spinor field $ \Psi(x)\quad (j=1/2)$ can be
written as
\begin{equation}
\begin{array}{l}
\widetilde{\Psi}(\widetilde{x})=R(a)\,\Psi(x), \\
g^{\mu}(\widetilde{x})\nabla_{\mu}\,\widetilde{\Psi}(\widetilde{x})=S(F)\,R(a)\,
\gamma^{l}D_{l}\,\Psi(x),
\end{array}
\label{R23}
\end{equation}
where
\begin{equation}
\begin{array}{l}
\nabla_{\mu}=\widetilde{\partial}_{\mu}+\Gamma_{\mu},\\
\Gamma_{\mu}(\widetilde{x})=(1/2)\,\Sigma^{\alpha\beta}\,V^{\nu}_{\alpha}(\widetilde{x})
\widetilde{\partial}_{\mu}\,V_{\beta\nu}(\widetilde{x}), \\
\widetilde{R}=\gamma^{0}R\gamma^{0},$  $ \quad \Sigma^{\alpha\beta}=
\frac{1}{4}[\gamma^{\alpha},\gamma^{\beta}],\\
\Gamma_{\mu}(\widetilde{x})=\frac{1}{4}
\Delta_{\mu,\alpha\beta}(\widetilde{x})\gamma^{\alpha}\gamma^{\beta},
\end{array}
\label{RG4}
\end{equation}
$ \Delta_{\mu,\alpha\beta}(\widetilde{x}) $ denote the Ricci
rotation coefficients. The unitary map matrix $R(a)$ is in the form
of Eq.~(\ref{R21}), provided we make single change
\begin{equation}
\begin{array}{l}
\Theta_{g}(\widetilde{x})=\frac{1}{2}\,\int_{0}^{\widetilde{x}}\,
R^{+}\left\{ g^{\mu}\Gamma_{\mu}R,\,
g_{\nu}\,d\,\widetilde{x}{}^{\nu}\right\}.
\end{array}
\label{R25}
\end{equation}
The calculations now give
\begin{equation}
\begin{array}{l}
S(F)=\frac{1}{8K}\psi^{l}_{\mu}\left\{ \widetilde{R}^{+}g^{\mu}R,\,
\gamma_{l}\right\}=inv, \\
K=\widetilde{R}^{+}\,R=\widetilde{R}^{+}_{g}\,R_{g},
\end{array}
\label{RG5}
\end{equation}
and hence
\begin{equation}
\begin{array}{l}
K=\exp\left( - \frac{1}{2}\,\int_{0}^{\widetilde{x}}\left( \left\{
R^{+} \widetilde{\Gamma}^{+}_{\mu}g^{\mu},\,
g_{\nu}d\widetilde{x}{}^{\nu}\right.\right.\right.\}R+
\\\left.\left. R^{+}\left\{ g^{\mu}\Gamma_{\mu}R,
\,g_{\nu}d\widetilde{x}{}^{\nu}\right\}\right)\right),
\end{array}
\label{R27}
\end{equation}
where $ \widetilde{\Gamma}^{+}_{\mu}=
\gamma^{0}\Gamma^{+}_{\mu}\gamma^{0}.$ Taking into account a
commutation relation $[R, g_{\nu}]=0$, and substituting~\citep{Foc}:
\begin{equation}
\begin{array}{l}
\widetilde{\Gamma}^{+}_{\mu}g^{\nu} + g^{\nu}\Gamma_{\mu}=
-\nabla_{\mu}g^{\nu}=0,
\end{array}
\label{RG6}
\end{equation}
 we obtain $ K = 1. $ Note that
\begin{equation}
\begin{array}{l}
\widetilde{U}_{R}^{+}U_{R}=\widetilde{R}U^{+} \widetilde{R'}^{+}R'U
R^{+}= \widetilde{R}U^{+} \widetilde{R}^{+}RU R^{+}, \end{array}
\label{RG7}
\end{equation}
and $\widetilde{R'}^{+}R'=\widetilde{R}^{+}R=1$, therefore
\begin{equation}
\begin{array}{l}
\widetilde{U}_{R}^{+}U_{R}=\gamma^{0}U_{R}^{+}\gamma^{0}U_{R}=1.
\end{array}
\label{RG8}
\end{equation}

\subsection{The GGP for any spin} \label{anysp}
The results above can be readily extended to any spin $j$. In doing
this we employ the Bargman-Wigner's wave functions for higher-spin
in flat space~\citep{Barg}. The formalism of GGP  will then hold,
wheras the field of arbitrary spin $j$ can be treated as a system of
2$j$ fermions of half-integer spin. The wave function of
spin-$j=n/2$ particle with momentum ${\vec p}$ defined on the
$M_{4}$ can be obtained by Lorentz transformation from the symmetric
Dirac spinor of rank $n$ corresponding to the particle in the rest
$U_{\alpha_{1}\ldots\alpha_{n}}(0)$ implying
$(\gamma_{4}-1)^{\beta'}_{\beta} U_{\beta'\beta_{2}\ldots
\beta_{n}}(0)$ for each index
\begin{equation}
\begin{array}{l}
U_{\beta_{1}\ldots\beta_{n}}({\vec p})=
S^{\beta'_{1}}_{\beta_{1}}(\alpha({\vec p})) \ldots
S^{\beta'_{n}}_{\beta_{n}}(\alpha({\vec p}))
U_{\beta'_{1}\ldots\beta'_{n}}(0), \end{array}\label{A1}
\end{equation}
where
\begin{equation}
\begin{array}{l}
\overline{U}'({\vec p})\,U'({\vec p})=\overline{U}(\Lambda^{-1}{\vec
p})\, U(\Lambda^{-1}{\vec p}).
\end{array}
\label{RG10}
\end{equation}
A spin part is written $
\Sigma_{\mu\nu}=(1/2)\sum^{n}_{r=1}\Sigma^{(r)}_{\mu\nu}, $ where a
matrix $\Sigma^{(r)}_{\mu\nu}$ acts only on the r-th index
\begin{equation}
\begin{array}{l}
\left(\Sigma^{(r)}_{\mu\nu}\right)^{\beta_{1}\ldots\beta_{n}}
_{\alpha_{1}\ldots\alpha_{n}}=\delta^{\beta_{1}}_{\alpha_{1}}\cdots
\delta^{\beta_{r-1}}_{\alpha_{r-1}}
\left(\Sigma^{(r)}_{\mu\nu}\right)^{\beta_{r}}_{\alpha_{r}}
\delta^{\beta_{r+1}}_{\alpha_{r+1}}\cdots
\delta^{\beta_{n}}_{\alpha_{n}},
\end{array}
\label{RG11}
\end{equation}
that $ \Sigma^{(r)}_{\mu\nu}=(1/4)[\gamma^{r}_{\mu},
\,\gamma^{r}_{\nu}], $ and
\begin{equation}
\begin{array}{l}
\left(\gamma^{r}_{\mu}\right)^{\beta_{1}\ldots\beta_{n}}
_{\alpha_{1}\ldots\alpha_{n}}=\delta^{\beta_{1}}_{\alpha_{1}}\cdots
\delta^{\beta_{r-1}}_{\alpha_{r-1}}
\left(\gamma^{r}_{\mu}\right)^{\beta_{r}}_{\alpha_{r}}
\delta^{\beta_{r+1}}_{\alpha_{r+1}}\cdots
\delta^{\beta_{n}}_{\alpha_{n}}.
\end{array}
\label{RG12}
\end{equation}
The spin-$j$ field $\Phi_{\beta_{1}\ldots\beta_{n}}(x)$
(Eq.~(\ref{A1})) takes values in standard fiber over $
x:\pi^{-1}(\mathcal{U}_{i})=\mathcal{U}_{i}\times \mathbb{F}_{x}. $
In the framework of GGP, the mapped spin-$j$ field
$\widetilde{\Phi}{}^{(r)}_{\beta_{1}\ldots\beta_{n}}(\widetilde{x}),$
where $ \widetilde{\Phi}{}^{(r)}=R^{(r)}\,\Phi, $ takes values in
the fiber over $
\widetilde{x}:\widetilde{\pi}{}^{-1}(\widetilde{\mathcal{U}}_{i})=\widetilde{\mathcal{U}}_{i}\times
\widetilde{\mathbb{F}}_{\widetilde{x}}.$ The unitary map matrix
$R^{(r)}$ is given in the form of Eq.~(\ref{R21}) and
Eq.~(\ref{R25}), but now refers to the $r$-th index. The Lagrangian
of this field will be invariant under the local gauge
transformations
\begin{equation}
\begin{array}{l}
\widetilde{\Phi}'(\widetilde{x})=U_{R}^{(r)}\,\widetilde{\Phi}
(\widetilde{x}),
\\ \left( g^{\mu}_{(r)}(\widetilde{x})\,\nabla_{\mu}^{(r)}\widetilde{\Phi} (\widetilde{x})\right)' =
U_{R}^{(r)} \left(
g^{\mu}_{(r)}(\widetilde{x})\,\nabla_{\mu}^{(r)}\widetilde{\Phi}
(\widetilde{x})\right),
\end{array}
\label{A2}
\end{equation}
where $ g^{\mu}_{(r)}(\widetilde{x})=
V^{\mu}_{\alpha}(\widetilde{x})\gamma^{\alpha}_{(r)},$
$\nabla_{\mu}^{(r)}$ denotes the covariant derivative on $R_{4}$
defined only for the $r$-th index by the conventional substitution
~\citep{Bir}:
\begin{equation}
\begin{array}{l}
\nabla_{\mu}^{(r)}\widetilde{U}_{\beta_{1}\ldots\beta_{n}}
\rightarrow \\\Lambda^{\alpha'}_{\alpha}(\widetilde{x})
S^{\beta'_{1}}_{\beta_{1}}(\alpha(\Lambda)) \ldots
S^{\beta'_{n}}_{\beta_{n}}(\alpha(\Lambda))
\nabla_{\mu}^{(r)}\widetilde{U}_{\beta_{1}\ldots\beta_{n}}.
\end{array}
\label{A3}
\end{equation}
Here, as usual, we denote $
\nabla_{\alpha}^{(r)}=V^{\mu}_{\alpha}\,(\widetilde{\partial}_{\mu}+\Gamma_{\mu}^{(r)}),
$ and that
\begin{equation}
\begin{array}{l}
\Gamma_{\mu}^{(r)}=\frac{1}{2}
\,\Sigma^{\alpha\beta}_{(r)}\,V^{\nu}_{\alpha}(\widetilde{x})
\,\partial_{\mu}V_{\beta\nu}(\widetilde{x})=\frac{1}{4}
\Delta_{\mu,\alpha\beta}^{(r)}\,\gamma^{\alpha}_{(r)}\,\gamma^{\beta}_{(r)},
\end{array}
\label{A4}
\end{equation}
$\Delta^{(r)}_{\mu,\alpha\beta}(\widetilde{x})$ are the Ricci
rotation coefficients. The Eqs.~(\ref{A2}) hold if
\begin{equation}
\begin{array}{l}
g^{\mu}_{(r)}\,\nabla_{\mu}^{(r)}\widetilde{\Phi}{}^{(r)}=
R^{(r)}\,S^{(r)}\,(\gamma^{l} \,D_{l}\,\Phi),
\end{array}
\label{A5}
\end{equation}
 where $D_{l}=\partial_{l}-
ig\, a_{l}(x)$, the $R^{(r)},\,S^{(r)}(F)$ are, respectively,
unitary map matrix and gauge invariant scalar function of
Eq.~(\ref{R23}) but referred to $r$-th index. According to the
results of subsection A.2, we get
\begin{equation}
\begin{array}{l}
\widetilde{U^{(r)}_{R}}^{+}\,U^{(r)}_{R}=\gamma^{0}\,
{U^{(r)}_{R}}^{+}\,\gamma^{0}\,U^{(r)}_{R}=1.
\end{array}
\label{RG12}
\end{equation}
The Lagrangian of the spin-$j$ field may be recast as
\begin{equation}
\begin{array}{l}
L(x)=J_{\psi} \widetilde{L}(\widetilde{x})= \\J_{\psi} \left\{
\frac{i}{2}\left[
\overline{\widetilde{\Phi}}{}^{(r)}(\widetilde{x})\,g_{(r)}^{\mu}(\widetilde{x})
\nabla^{(r)}_{\mu}\widetilde{\Phi}{}^{(r)}(\widetilde{x})-
\right.\right. \\\left.\left.
(\nabla^{(r)}_{\mu}\overline{\widetilde{\Phi}}{}^{(r)}(\widetilde{x}))\,
g^{\mu}_{(r)}(\widetilde{x}) \widetilde{\Phi}{}^{(r)}(\widetilde{x})
\right]- \right. \\\left.
m\overline{\widetilde{\Phi}}{}^{(r)}(\widetilde{x})\widetilde{\Phi}{}^{(r)}
(\widetilde{x}) \right\}=  \\J_{\psi} \left\{ S^{(r)}(a) \frac{i}{2}
\left[\overline{\Phi}\,\gamma^{l}_{(r)}D_{l}\,\Phi-
(D_{l}\overline{\Phi})\,\gamma^{l}_{(r)}\Phi\right]-
m\overline{\Phi}\,\Phi\right\},
\end{array}
\label{A6}
\end{equation}
where $J_{\psi}\equiv \| \psi\|\,\sqrt{-g}$. Generalized
Bargman-Wigner's equation for the spin-$j=\frac{n}{2}$ particle in
curved space stems from the Lagrangian eq.~(\ref{A6}):
\begin{equation}
\begin{array}{l}
\left({g'}^{\mu} \,\nabla'_{\mu}\widetilde{\Phi}^{(r)}-m\right)
^{\beta'}_{\beta}
\,\widetilde{\Phi}_{\beta'\beta_{2}\ldots\beta_{n}}
(\widetilde{{\vec p}})=  \\\left[ R'\left(S'\,D-m\right)\right]
^{\beta'}_{\beta} \,\Phi_{\beta'\beta_{2}\ldots\beta_{n}}({\vec
p})=0,
\end{array}
\label{A7}
\end{equation}
where $R',S',\ldots$ refer to index $\beta'$.

\subsection{The conserved currents}
The general transformation of the group $G_{D}$ is written as
\begin{equation}
\begin{array}{l}
G_{D}=\overline{Q}(\overline{a})H(\theta),
\end{array}
\label{R640}
\end{equation}
where the transformation $\overline{Q}(\overline{a})$  belongs to
the left adjacent class $G_{D}/H$. Acting from the left by the group
element $g$ we define the transformations of parameters
$\overline{a}$ and $\theta$:
\begin{equation}
\begin{array}{l}
G_{D}(g)\overline{Q}(\overline{a})H(\theta)=\overline{Q}(\overline{a}'
(\overline{a},\,g))H(\theta'(\theta,\, \overline{a},\, g )).
\end{array}
\label{R641}
\end{equation}
The Cartan's forms allow an exponentiation of the group element
$\overline{Q}(a)=\exp (i \overline{a}{}^{i}K_{i})$, which determines
the modified distortion fields $\overline{a}(a)$
\begin{equation}
\begin{array}{l}
d\,\overline{e}_{A}(\overline{a})= \left[\exp (-i
\overline{a}{}^{i}K_{i})\,d\,\exp (i \overline{a}{}^{i}K_{i})\right]
\,\overline{e}_{A}(\overline{a}).
\end{array}
\label{R642}
\end{equation}
Exponential parametrization of the finite transformations of the
group is equivalent to the choice of normal coordinates in quotien
space $G_{D}/H$. The solution of the structure
equations~(\ref{R647}) reads
\begin{equation}
\begin{array}{l}
\omega^{i}(\overline{a},\,d\,\overline{a})=\left(
\sin\,\sqrt{\tau}/\sqrt{\tau}\right)^{i}_{k}\,d\,\overline{a}{}^{k}, \\
\vartheta^{\alpha}(\overline{a},\,d\,\overline{a})=\left[\left(1-
\cos\,\sqrt{ \tau}\right)/
\tau\right]^{i}_{k}\,d\,\overline{a}{}^{k}\,\varepsilon_{\alpha i
l}\,\overline{a}{}^{l},\\
\tau^{i}_{k}=-\varepsilon_{ij\alpha}\varepsilon_{\alpha k
l}\,\overline{a}{}^{j}\,\overline{a}{}^{k}.
\end{array}
\label{R747}
\end{equation}
By virtue of the relations
\begin{equation}
\begin{array}{l}
\left( \tau^{(n)}\right)^{i}_{j}=(\overline{a}{}^{2})^{n-1}\left(
\tau^{(1)}\right)^{i}_{j}, \\
\tau^{i}_{j}=\varepsilon_{ilk}\varepsilon_{kjn}\overline{a}{}^{l}\overline{a}{}^{n}=
\overline{a}{}^{2}\delta_{ij}-\overline{a}{}^{i}\overline{a}{}^{j},
\end{array}
\label{R748}
\end{equation}
the Eq.~(\ref{R747}) may be recast as
\begin{equation}
\begin{array}{l}
\omega^{i}_{A}=\omega^{i}(\overline{a},\,\partial_{A}\,\overline{a})=
\partial_{A}\,\overline{a}{}^{i}+\\
(\delta_{ik}-\overline{a}{}^{i}\overline{a}{}^{k}/\overline{a}{}^{2})\,
\left(
\sin\,\sqrt{\overline{a}{}^{2}}/\sqrt{\overline{a}{}^{2}}-1\right)\partial_{A}\,\overline{a}{}^{k},
\\
\vartheta^{i}_{A}=\vartheta^{i}(\overline{a},\,\partial_{A}\,\overline{a})=\left[\left(1-
\cos\,\sqrt{ \overline{a}{}^{2}}\right)/
\overline{a}{}^{2}\right]\partial_{A}\,\overline{a}{}^{l}\,\varepsilon_{i
lj}\,\overline{a}{}^{j}.
\end{array}
\label{R749}
\end{equation}
In exponential parametrization the Eq.~(\ref{R641}) gives
\begin{equation}
\begin{array}{l}
G_{D}(g)\exp (i \overline{a}{}^{i}K_{i})\exp (i
\theta^{\alpha}I_{\alpha})= \\\exp
\left[i\overline{a}{}^{i}(\overline{a},\,g)K_{i}\right]\exp \left[i
\theta'{}^{\alpha}(\overline{a},\,g)I_{\alpha}\right].
\end{array}
\label{R844}
\end{equation}
According to the Eq.~(\ref{R844}), the transformation of parameters
$\overline{a}$ and $\theta$ at the infinitesimal translation
$G_{D}(g)=(1+i\varepsilon^{i}\,K_{i})+O(\varepsilon^{2})$ can be
written as
\begin{equation}
\begin{array}{l}
(1+i\varepsilon^{i}\,K_{i})\exp (i K_{i}\overline{a}{}^{i})= \\\exp
\left[iK_{i}(\overline{a}{}^{i}+
\delta\,\overline{a}{}^{i}(\overline{a},\,\varepsilon))\right] \exp
\left[i
\delta\,\theta^{\alpha}(\overline{a},\,\varepsilon))I_{\alpha}\right].
\end{array}
\label{R801}
\end{equation}
Employing the Feynman's method of ordering by means of auxiliary
parameter ($t$)~\citep{Gel} which implies
$\int^{y}_{x}\,A(t)d\,t=A(y-x)$, in standard manner we expand
\begin{equation}
\begin{array}{l}
\exp \left[iK_{i}(\overline{a}{}^{i}+
\delta\,\overline{a}{}^{i})\right] = \exp
(iK_{i}\overline{a}{}^{i})+i\exp (iK_{i}\overline{a}{}^{i})\times\\
\times\int^{1}_{0}\,d\,t\,\exp
(-iK_{i}\overline{a}{}^{i}\,t)\,(\delta\,\overline{a}{}^{i}\,K_{i})
\,\exp (iK_{i}\overline{a}{}^{i}\,t) + \cdots.
\end{array}
\label{R808}
\end{equation}
Then the Eq.~(\ref{R801}) gives
\begin{equation}
\begin{array}{l}
i\varepsilon^{i}\,K_{i}(1)=i\int^{1}_{0}\,d\,t
\delta\,\overline{a}{}^{i}K_{i}(t)+i
\delta\,\theta^{\alpha}(\overline{a},\,\theta)I_{\alpha},
\end{array}
\label{R802}
\end{equation}
where
\begin{equation}
\begin{array}{l}
K_{j}(t)=\exp (-i K_{}\overline{a}{}^{i}t)K_{j}\exp (i
K_{i}\overline{a}{}^{i}t).
\end{array}
\label{R803}
\end{equation}
In analogy, defining the $I_{\alpha}(t)$, and taking derivatives of
the $K_{j}(t)$  and $I_{\alpha}(t)$ with respect to parameter ($t$),
we obtain
\begin{equation}
\begin{array}{l}
\frac{\partial}{\partial t}\,K_{j}(t)=i\varepsilon_{\alpha j
i}\overline{a}{}^{i}\,I_{\alpha}(t); \quad K_{j}(0)=K_{j},\\
\frac{\partial}{\partial t}\,I_{\alpha}(t)=i\varepsilon_{l\alpha k
}\overline{a}{}^{k}\,K_{l}(t); \quad I_{\alpha}(0)=0,
\end{array}
\label{R804}
\end{equation}
the solution of which can be written in the form of Eq.~(\ref{R747})
\begin{equation}
\begin{array}{l}
K_{j}(t)=\left( \cos\,\sqrt{\tau}\,t\right)^{i}_{j}\,K_{j}+
\\i\left(
\sin\,\sqrt{\tau}\,t/\sqrt{\tau}\right)^{l}_{j}\,\varepsilon_{\alpha
lk }\,\overline{a}{}^{k}\,I_{\alpha}.
\end{array}
\label{R805}
\end{equation}
Inserting this solution into the Eq.~(\ref{R802}) and equating the
coefficients at the same generators $K_{j}I_{\alpha}$, we finally
obtain
\begin{equation}
\begin{array}{l}
\delta\,\overline{a}{}^{i}(\overline{a},\,\theta))=-(\sqrt{\tau}\,\coth
\,\sqrt{\tau})^{i}_{k}\varepsilon^{k}+O(\varepsilon^{2}),\\
\delta\,\theta^{\alpha}(\overline{a},\,\theta))=\{
[\sin\,\sqrt{\tau}-
\coth \,\sqrt{\tau}(1-\\
\cos\,\sqrt{\tau})]/\sqrt{\tau}\}^{l}_{i}\,\varepsilon^{i}\,\varepsilon_{\alpha
lk}\overline{a}{}^{k}.
\end{array}
\label{R806}
\end{equation}
The transformation of parameters in case of rotation
$G_{D}(g')=(1+i\xi^{\alpha}\,I_{\alpha})+O(\xi^{2})$ can be obtained
in the same manner as
\begin{equation}
\begin{array}{l}
\delta\,\overline{a}{}^{i}(\xi))=-\varepsilon_{i\alpha
k}\,\xi^{\alpha}\,\overline{a}{}^{k}+O(\xi^{2}),\quad
\delta\,\theta^{\alpha}(\overline{a},\,\xi))=\xi^{\alpha}.
\end{array}
\label{R807}
\end{equation}
Let the fields undergo the infinitesimal transformations
\begin{equation}
\begin{array}{l}
\Phi^{j}(\eta)\rightarrow \Phi^{j}(\eta)+
\varepsilon(\eta)^{k}\,\prod^{j}_{k}(\Phi),
\end{array}
\label{R845}
\end{equation}
where $\prod^{j}_{k}(\Phi)$ is the nonlinear function of the fields.
The current related to these transformations
\begin{equation}
\begin{array}{l}
J^{A}_{k}=-\delta\,L(\Phi,\,
\partial\,\Phi)/\delta\,(\partial_{A}\,\varepsilon^{k})=\\-\prod^{j}_{k}(\Phi)\,
\delta\,L(\Phi,\,
\partial\,\Phi)/\delta\,(\partial_{A}\,\Phi^{j}),
\end{array}
\label{R846}
\end{equation}
implies
$$
\begin{array}{l}
\partial_{A}\,J^{A}_{k}=-\delta\,L(\Phi,\,
\partial\,\Phi)/\delta\,\varepsilon^{k}.
\end{array}
$$
If the Lagrangian is invariant with respect to constant
transformations of Eq.~(\ref{R845}), then the corresponding currents
are conserved. According to equations~(\ref{R747}), ~(\ref{R806})
and~(\ref{R807}) the conserved currents can be calculated in normal
coordinates as
\begin{equation}
\begin{array}{l}
J^{k}(\overline{a},\,\partial \,\overline{a})\equiv
\omega^{k}(2\overline{a},\,\partial \,\overline{a})= (\sin
\,2\sqrt{\tau}/2\sqrt{\tau})^{k}_{i}\partial\,\overline{a}{}^{i},
\\
J^{\alpha}(\overline{a},\,\partial \,\overline{a})\equiv
\vartheta^{\alpha}(2\overline{a},\,\partial \,\overline{a})=\\
-\left[\left(1- \cos\,2\sqrt{ \tau}\right)/
2\tau\right]^{i}_{k}\,\partial\,\overline{a}{}^{k}\,\varepsilon_{\alpha
i l}\,\overline{a}{}^{l},
\end{array}
\label{R848}
\end{equation}
where we left the indices $(A)$ implicit.

\end{document}